\documentclass[onecolumn]{IEEEtran}
\usepackage{amsmath, amssymb}
\usepackage{graphicx}
\usepackage{bm}

\newtheorem{theorem}    {Theorem}
\newtheorem{lemma}{Lemma}
\newtheorem{corollary}{Corollary}
\newtheorem{remark}{Remark}
\newtheorem{example}{Example}

\newcommand{\Tvec}{\vec{\bm{T}}}
\newcommand{\Mvec}{\vec{\bm{M}}}
\newcommand{\rhovec}{\vec{\bm{\rho}}}
\newcommand{\sigmavec}{\vec{\bm{\sigma}}}
\newcommand{\Phivec}{\vec{\bm{\Phi}}}
\newcommand{\Svec}{\vec{\bm{S}}}

\newcommand{\defeq}{\stackrel{\rm def}{=}}
\newcommand{\bR}{\mathbb{R}}
\newcommand{\bC}{\mathbb{C}}
\newcommand{\cH}{{\cal H}}
\newcommand{\cX}{{\cal X}}
\newcommand{\cXn}{{\cal X}^n}
\newcommand{\Xn}{X^n}
\newcommand{\bX}{{\bf X}}
\newcommand{\cYn}{{\cal Y}_n}
\newcommand{\cT}{{\cal T}}
\newcommand{\cTn}{{\cal T}_n}
\newcommand{\Tr}{{\rm Tr}\,}
\newcommand{\cHn}{{\cal H}_n}
\newcommand{\cA}{{\cal A}}
\newcommand{\cAn}{{\cal A}_n}
\newcommand{\cB}{{\cal B}}

\newcommand{\Tn}{T_n}

\newcommand{\rhon}{\rho_n}
\newcommand{\sigman}{\sigma_n}

\newcommand{\alphan}{\alpha_n}
\newcommand{\betan}{\beta_n}
\newcommand{\betancomp}{\beta_n^{\,c}}
\newcommand{\etan}{\eta_n}
\newcommand{\zetan}{\zeta_n}
\newcommand{\etancomp}{\eta_n^{\,c}}
\newcommand{\zetancomp}{\zeta_n^{\,c}}
\newcommand{\alphasup}{\overline{\alpha}}
\newcommand{\alphainf}{\underline{\alpha}}
\newcommand{\betasup}{\overline{\beta}}
\newcommand{\betainf}{\underline{\beta}}
\newcommand{\etasup}{\overline{\eta}}
\newcommand{\etainf}{\underline{\eta}}
\newcommand{\etasupcomp}{\overline{\eta}^{\,c}}
\newcommand{\zetasup}{\overline{\zeta}}
\newcommand{\zetainf}{\underline{\zeta}}
\newcommand{\zetasupcomp}{\overline{\zeta}^{\,c}}
\newcommand{\zetainfcomp}{\underline{\zeta}^c}
\newcommand{\sigmainf}{\underline{\sigma}}
\newcommand{\sigmasupast}{\overline{\sigma}^{\,*}}

\newcommand{\Phin}{\Phi_n}
\newcommand{\Fn}{F_n}
\newcommand{\Gn}{G_n}
\newcommand{\gamman}{\gamma_n}
\newcommand{\psin}{\psi_n}
\newcommand{\varphin}{\varphi_n}
\newcommand{\deltan}{\delta_n}

\makeatletter
\newcommand{\lleq}{\mathrel{\mathpalette\gl@align<}}
\newcommand{\ggeq}{\mathrel{\mathpalette\gl@align>}}
\newcommand{\gl@align}[2]{
\vbox{\baselineskip\z@skip\lineskip\z@
\ialign{$\m@th#1\hfil##\hfil$\crcr#2\crcr{}_{{}_{(=)}}\crcr}}}
\makeatother

\newcommand{\supD}{\overline{\specD}}
\newcommand{\infD}{\underline{\specD}}
\newcommand{\specD}{D}
\newcommand{\supH}{\overline{\specH}}
\newcommand{\infH}{\underline{\specH}}
\newcommand{\specH}{H}

\newcommand{\cHtensor}{{\cal H}^{\otimes n}}
\newcommand{\rhotensor}{\rho^{\otimes n}}
\newcommand{\sigmatensor}{\sigma^{\otimes n}}

\newcommand{\sepand}{\;\;\hbox{and}\;\;}

\newcommand{\altXn}{Y^n}
\newcommand{\altbX}{{\bf Y}}

\begin{document}
\title{An Information-Spectrum Approach to \\
Classical and Quantum Hypothesis Testing \\
for Simple Hypotheses}
\author{Hiroshi Nagaoka
\thanks{%
Hiroshi Nagaoka is with Graduate School of Information Systems,
The University of Electro-Communications, 
1-5-1, Chofugaoka, Chofu-shi, Tokyo, 182-8585, Japan 
(e-mail: nagaoka@is.uec.ac.jp).}
\and
Masahito Hayashi
\thanks{%
Masahito Hayashi was with the Laboratory for Mathematical Neuroscience, 
Brain Science Institute, RIKEN, 2--1 Hirosawa, Wako, Saitama, 351--0198, 
Japan 
(e-mail: masahito@brain.riken.go.jp). 
He is now with ERATO-SORST Quantum Computation and Information Project, 
Japan Science and Technology Agency (JST), 
201 Daini Hongo White Bldg. 5-28-3, Hongo, Bunkyo-ku, Tokyo, 113-0033, 
Japan (e-mail: masahito@qci.jst.go.jp), and 
Superrobust Computation Project, Information Science and Technology 
Strategic Core (21st Century COE by MEXT), 
Graduate School of Information Science and Technology, 
The University of Tokyo, 7-3-1, Hongo, Bunkyo-ku, Tokyo, 113-0033, Japan.
}}

\date{}
\maketitle

\begin{abstract}
The information-spectrum analysis made by Han for classical 
hypothesis testing for simple hypotheses is 
extended to a unifying framework 
including both classical and quantum hypothesis testing. 
The results are also applied to fixed-length source coding 
when loosening 
the normalizing condition for probability distributions
and for quantum states. 
We establish general formulas for several quantities relating to the 
asymptotic optimality of tests/codes in terms 
of classical and quantum information spectra. 
\end{abstract} 

\begin{keywords}
Information spectrum,
Quantum hypothesis testing,
Classical hypothesis testing,
Fixed-length source coding,
Optimal exponent
\end{keywords}

\section{Introduction}
One of the principal aims of information theory is 
to establish a link between two different kinds of quantities. 
One is an operational quantity which is defined as the optimal or limiting 
value of a concrete parameter such as code length, compression 
rate, transmission rate, convergence rate of error probabilities, etc.  
The other is an information quantity such as the entropy, 
divergence, mutual information, etc.  
Note that the latter, in its definition, is more abstract than the former, 
and the meaning of the latter is usually clarified by linking it to the 
former. 
In the so-called 
information spectrum method which first appeared in 
a series of joint papers of Han and Verd\'{u} 
(e.g., \cite{HanVerdu_output, VerduHan_general}), 
 the process of establishing such a link 
 is intentionally divided into 
two parts by introducing a third kind of quantity --- 
{\em information spectrum}, putting it between an
operational quantity and an abstract information quantity.  
This setting allows us to pursue many problems 
of information theory in their most general forms; see 
\cite{Han_book} for the whole perspective of the method.

For instance, let us consider two 
sequences of random variables 
$\bX = \{\Xn\}_{n=1}^\infty$ and 
$\altbX = \{\altXn\}_{n=1}^\infty$, 
where 
 $\Xn$ and $\altXn$ for each $n$ are supposed to take 
values in a common discrete (finite or countable)\footnote{
In this paper 
we only treat the discrete case to simplify the description 
when considering the classical hypothesis testing, 
although it is straightforward as pointed out in \cite{Han_book, 
Han_test} 
to extend the argument 
to the general case where $\{\cXn\}$ are arbitrary 
measurable spaces. }
set ${\cX}^n$ subject to 
probability distributions (mass functions)  
$P_{\Xn}$ and $P_{\altXn}$ respectively. 
Note that  $\cXn$ does not need to be 
the product set $\cX\times \cdots 
\times \cX$ of an $\cX$, although 
the notation suggests that the product set 
is a representative example of $\cXn$. 
Han \cite{Han_book, Han_test} studied 
the hypothesis testing problem for the simple 
hypotheses consisting of the general 
processes $\bX$ and $\altbX$ 
by means of 
the information spectrum, which is 
the asymptotic behavior of the random variable 
$\frac{1}{n}\log \frac{P_{\Xn}(\Xn)}{P_{\altXn} (\Xn)}$ 
(or $\frac{1}{n}\log 
\frac{P_{\Xn}(\altXn)}{P_{\altXn} (\altXn)}$)
in this case.  
He succeeded in representing several asymptotic characteristics of 
hypothesis testing  in terms of the information spectrum with no or 
very few assumptions 
on the processes.  The term `spectrum' is intended to 
mean that the scope of the theory covers the general 
case when the probability distribution of 
$\frac{1}{n}\log \frac{P_{\Xn}(\Xn)}{P_{\altXn} (\Xn)}$ 
does not necessarily get concentrated 
at a point, but may spread out, 
as $n\rightarrow\infty$. 

The purpose of the present paper is to 
extend, complement and refine Han's analysis of 
hypothesis testing from several viewpoints.  
The biggest motivation comes from the question of 
how to extend the analysis to 
quantum hypothesis 
testing.  Following the above setting, we 
are naturally led to consider the problem of 
hypothesis testing for the simple hypotheses 
consisting of two sequences of quantum states 
$\rhovec = \{\rhon\}_{n=1}^\infty$ and 
$\sigmavec = \{\sigman\}_{n=1}^\infty$, 
where $\rhon$ and $\sigman$ are density operators 
on a common Hilbert space $\cHn$ for each $n$.  
However, it is by no means obvious whether 
a similar analysis to that of Han is applicable to the 
quantum setting.  
We show in this paper that it is actually possible to extend Han's results 
by appropriately choosing a quantum 
analogue of the information spectrum 
so that both the classical and quantum cases are 
treated in a unifying framework. 
Although this does not mean that application to a special class 
of quantum processes such as i.i.d.\ (independent and identically 
distributed) ones immediately yields significant results, it seems 
to suggest a new approach 
to studying the quantum asymptotics and 
to elucidating a general principle underlying 
classical/quantum information theory. 

It should be noted that, even though the statements of our theorems are 
almost 
parallel to those for the classical setting, 
some of the proofs are essentially 
different from the original proofs of Han. 
The technique of information-spectrum slicing, which 
was effectively used in \cite{Han_book, Han_test, Han-distortion} 
to prove several important theorems, consists of 
a procedure of partitioning  
a set and does not straightforwardly apply  
to the quantum setting.   We are thus forced to 
look for another idea for proofs.  Fortunately, we have successfully 
found a way which does not need information-spectrum slicing and 
is applicable to the quantum setting.   Moreover, the new proofs 
are much simpler than the original ones even in the classical case.  
This simplification is a byproduct of our attempt to pursue quantum 
extensions.  

This paper also contains 
results such as those of 
Theorems~\ref{thm:Be*} and \ref{thm:R*e}  
which improve the corresponding original theorems 
when applied to the classical setting.  In addition, 
from the beginning, we treat generalized hypothesis testing 
in the sense of Han \cite{Han_book, Han_test}, 
namely that the 
alternative hypothesis $P_{\altXn}$ can be  
any nonnegative measure.  This enables us to 
unify hypothesis testing and fixed-length source coding 
in a natural way. 

This paper aims at presenting a unifying framework 
to treat the classical and quantum generalized hypothesis testing problem 
in the most general and simplest manner.   After 
presenting the notation 
in section~\ref{sec:preliminaries}, the concept of information spectrum 
is introduced in section~\ref{sec:info_spec} for both classical and 
quantum cases.  In sections \ref{sec:stein}, \ref{sec:error_exp} and 
\ref{sec:correct_exp}, various types of asymptotic bounds on the 
hypothesis testing problems for classical and quantum general processes 
are studied,  basically following the problem settings and the notation 
given in  \cite{Han_book, Han_test}.  In section~\ref{sec:i.i.d.} we 
make some observations on Stein's lemma for classical and 
quantum i.i.d.\ processes 
in the light of the results of 
section~\ref{sec:stein}.
Applications to 
the classical fixed-length source coding are presented in 
section~\ref{sec:source}, and concluding remarks are 
given in section~\ref{sec:conclusion}. 

\section{A unifying description of 
classical and quantum generalized 
hypothesis testing}
\label{sec:preliminaries}

In this section we 
present a common language to 
treat classical and quantum hypothesis testing 
and fixed-length source coding 
in a unifying manner.  We 
begin by considering 
the classical case.  Suppose that we are given 
a sequence of discrete sets $\vec{\cX} = 
\{\cXn\}_{n=1}^\infty$, a sequence of probability 
measures $\rhovec = \{\rhon\}_{n=1}^\infty$ and 
a sequence of nonnegative (not necessarily 
probability) measures
$\sigmavec = \{\sigman\}_{n=1}^\infty$, 
which are represented by mass functions 
$\rhon : \cXn\rightarrow [0,1]$ with $\sum_{x\in\cXn} \rhon (x) 
= 1$ and $\sigman : \cXn\rightarrow [0,\infty)$. 
In the usual hypothesis testing problem, 
both $\rhon$ and $\sigman$ are probability 
measures  denoted as 
$\rhon = P_{\Xn}$ and $\sigman = P_{\altXn}$. 
On the other hand, $\sigman$ should be taken to be the counting measure on 
$\cXn$ when considering the source coding problem 
(see \cite{Han_book, Han_test} and section~\ref{sec:source} below).  
For a function (random variable) 
$A : \cXn\rightarrow\bR$, we write 
\[
\rhon[ A ] = \sum_{x\in\cXn} \rhon(x) \, A(x) 
\quad\mbox{and}\quad
\sigman[A] = \sum_{x\in\cXn} \sigman(x) \, A(x). 
\]

Let 
$\cTn$ be the set of $[0,1]$-valued functions defined on 
$\cXn$.  When both $\rhon$ and $\sigman$ are probability measures, 
we regard an element $\Tn$ of $\cTn$ as a randomized 
test for 
the  simple hypotheses $\{\rhon, \sigman\}$ by interpreting 
$\Tn (x)$  ($\in [0,1]$) as the probability of accepting the hypothesis 
$\rhon$ 
when the data $x$ is observed.  In particular, 
a deterministic test is an element of $\cTn$ 
taking values in $\{0,1\}$, which is the characteristic function of 
the acceptance region $\{x\in\cXn\,|\, \Tn(x) =1\}$ 
for the hypothesis $\rhon$. 
Depending on whether the true distribution is $\rhon$ or $\sigman$, 
the probability of accepting the hypothesis $\rhon$ turns out 
to be  $\rhon[\Tn]$ or $\sigman[\Tn]$, and 
the error probabilities of the first and 
second kinds are represented as 
\begin{equation}
\label{alphanTn}
\alphan[\Tn] \defeq  1-\rhon[\Tn]
\quad\mbox{and}\quad
\beta[\Tn] \defeq \sigman[\Tn].
\end{equation}
In the general situation where $\sigman$ is an 
arbitrary nonnegative measure, 
we still call elements of $\cTn$ {\em tests} and 
use the same notation as in  (\ref{alphanTn}). 
Letting $I$ and $0$ denote the constant functions on $\cXn$ 
such that $I(x)= 1$ and $0(x)= 0$ for all $x\in\cXn$, a test $\Tn$ is 
characterized as a function such that $0 \leq \Tn\leq I$, 
where, and in the sequel, we write $A\leq B$ for functions 
$A$ and $B$  when $A(x)\leq B(x)$ for all $x$.

Let us turn to the quantum case.   Suppose that a sequence of 
Hilbert spaces $\vec{\cH} = \{\cHn\}_{n=1}^\infty$ is given. 
Let $\rhovec = \{\rho_n\}_{n=1}^\infty$ be a sequence of 
density operators (i.e., nonnegative self-adjoint operators with trace one) 
on $\{\cHn\}$, 
and $\sigmavec = \{\sigma_n\}_{n=1}^\infty$ be a sequence of 
bounded nonnegative self-adjoint  
(but not necessarily density or trace-class)  operators on $\{\cHn\}$.  For 
a bounded self-adjoint operator $A$ on $\cHn$ we write 
\[ \rhon [A] = \Tr (\rhon A)  \quad{\rm and}\quad  
\sigman[A ] = \Tr (\sigman A) . 
\]
Let $\cTn$ be the 
set of self-adjoint operators 
$\Tn$ satisfying $0\leq\Tn\leq I$; i.e., both $\Tn$ and $I-\Tn$ 
are nonnegative with $I$ denoting the identity operator on $\cHn$.
An element $\Tn$ of $\cTn$ can be considered to 
represent a $\{0,1\}$-valued measurement on $\cHn$ by 
identifying it with the POVM (positive operator-valued 
measure) $\{\Tn (0), \Tn(1))\}= \{\Tn, I-\Tn\}$, and is 
called a {\em test} for the hypotheses $\{\rhon, \sigman\}$ 
with the interpretation that the measurement 
result $0$ means the acceptance of the hypothesis $\rhon$.

We define $\alphan[\Tn]$ and $\betan[\Tn]$ by the 
same equations as (\ref{alphanTn}), 
which turn out to be the error probabilities of the first and second 
kinds when both $\rhon$ and $\sigman$ are density operators. 

We have thus reached a common setting 
to treat 
generalized hypothesis testing of classical and 
quantum systems for which sequences 
\[ \rhovec = \{\rho_n\}_{n=1}^\infty, 
\quad  
\sigmavec = \{\sigma_n\}_{n=1}^\infty, 
\quad{\rm and}\quad
\vec{\cT} = \{\cT_n\}_{n=1}^\infty 
\]  
are given. Note that 
\begin{equation}
\label{cond:rhon}
 0=\rhon[0] \leq \rhon [\Tn] \leq \rhon [T_n'] \leq 
\rhon [I] = 1
\end{equation}
and 
\begin{equation}
\label{cond:sigman}
0=\sigman[0] \leq \sigman [\Tn] \leq \sigman [T_n'] \leq 
\sigman [I] \leq\infty
\end{equation}
always hold for any tests $\Tn, T_n'\in\cTn$ such that 
$\Tn\leq T_n'$. We shall work with this setting 
throughout this paper. 

\begin{remark}
\label{remark:*algebra}
{\rm 
Readers who are familiar with the language of operator algebras 
may immediately extend the setting to a more general one in which 
we are given a sequence of a certain kind of $*$-algebras 
$\vec{\cA}=\{\cAn\}_{n=1}^\infty$ containing 
the identity elements $I$, a sequence of states 
$\rhovec = \{\rho_n\}_{n=1}^\infty$  
(linear functionals mapping nonnegative elements to nonnegative 
numbers and the identity elements to $1$) and a sequence 
of positive weights $\sigmavec = \{\sigma_n\}_{n=1}^\infty$ 
(linear functionals mapping nonnegative elements to 
nonnegative numbers or $\infty$), with defining 
$\cTn = \{\Tn\in\cAn\,|\,  0\leq \Tn = T_n^{\,*}\leq I\}$. 
The classical case and 
the quantum case treated above correspond to 
$\cAn= L^\infty(\cXn)$ (the set of complex-valued bounded 
functions on $\cXn$) and $\cAn= \cB(\cHn)$ (the 
set of bounded operators on $\cHn$) respectively.  
}
\end{remark}

The following notation is introduced in order to
represent several variations of error exponents 
in a unifying manner. 
Given a sequence of tests $\Tvec =\{\Tn\}_{n=1}^\infty$ 
such that $\Tn\in\cTn$ ($\forall n$), let 
\begin{eqnarray*}
\etan [T_n] &\defeq 
- \frac{1}{n} \log \alpha_n[T_n] &= - \frac{1}{n} \log (1-\rho[T_n]), \\
\zetan [T_n] &\defeq
- \frac{1}{n} \log \beta_n[T_n] &= - \frac{1}{n} \log \sigma[T_n], 
\end{eqnarray*}
and 
\begin{eqnarray*}
\alphainf [\Tvec ] &\!\!\defeq 
\displaystyle
\liminf_{n\rightarrow\infty} 
\alphan [T_n], \qquad 
\alphasup [\Tvec ] &\!\!\defeq 
\limsup_{n\rightarrow\infty} 
\alphan [T_n], \\
\betainf [\Tvec ] &\!\!\defeq 
\displaystyle
\liminf_{n\rightarrow\infty} 
\betan [T_n], \qquad 
\betasup [\Tvec ] &\!\!\defeq 
\limsup_{n\rightarrow\infty} 
\betan [T_n], \\
\etainf [\Tvec ] &\!\!\defeq 
\displaystyle
\liminf_{n\rightarrow\infty} 
\etan [T_n], \qquad 
\etasup [\Tvec ] &\!\!\defeq 
\limsup_{n\rightarrow\infty} 
\etan [T_n], \\
\zetainf [\Tvec ] &\!\!\defeq  
\displaystyle
\liminf_{n\rightarrow\infty} 
\zetan [T_n] , \qquad 
 \zetasup [\Tvec ] &\!\!\defeq  
\limsup_{n\rightarrow\infty} 
\zetan [T_n] .
\end{eqnarray*}
When $\Tvec = \{\Tn\}$ is replaced with its {\em complement} 
$\Tvec^{\,c} = \{T_n^{\,c} \defeq I - \Tn\}$, 
we add the superscript 
${}^c$ to these symbols as $\alpha_n^{\,c}[\Tn] = \alphan [T_n^{\,c}]$, 
$\eta_n^{\,c} [\Tn] = \etan [T_n^{\,c}]$, $\zetainfcomp [\Tvec ] 
= \zetainf [{\Tvec}^{\,c}]$, etc.  

\section{Information spectrum and likelihood tests}
\label{sec:info_spec}

As mentioned in the introduction, the information spectrum 
for classical hypothesis testing is the asymptotic 
behavior of the random variable 
\[ Z_n \defeq \frac{1}{n} \log\frac{\rhon (\Xn)}{\sigman (\Xn)}, 
\]
where $\Xn$ is supposed to be subject to 
the probability distribution $\rhon$.  
Han \cite{Han_book, Han_test} called $Z_n$ the 
divergence-density rate and derived several formulas for 
representing the asymptotic characteristics of the classical 
hypothesis testing problem in terms of the information spectrum. 
Now we are led to the following question; what is the quantum 
analogue of the information spectrum?  
At a first glance, it may seem to be natural to consider the 
quantum observable represented by the self-adjoint 
operator $\frac{1}{n} (\log \rhon -\log\sigman)$ and 
its probability distribution under the quantum state $\rhon$. 
Unfortunately, this line is not directly linked to the 
hypothesis testing problem.  We give up seeking the quantum 
analogue of $Z_n$,  but instead seek that of a likelihood test 
$S_n(a) : \cXn \rightarrow [0,1]$ obeying 
\begin{equation}
\displaystyle
S_n(a) (x) = \left\{
\begin{array}{ccc}
1 &{\rm if}& \frac{1}{n} \log \frac{\rhon(x)}{\sigman(x)}  > a 
\\
0 &{\rm if}& \frac{1}{n} \log\frac{\rhon(x)}{\sigman(x)} < a 
\end{array}
\right.
\label{classical_S_n(a)}
\end{equation}
where $a$ is an arbitrary real number.  
Note that there is an ambiguity in this definition 
of $S_n(a)$ when some $x$ satisfies 
$\frac{1}{n} \log\frac{\rhon(x)}{\sigman(x)} = a $, 
 including two special cases where 
$S_n(a)$ are the deterministic tests with the 
acceptance regions 
\begin{align*}
&\Bigl\{x\in\cXn\;\Big|\; \frac{1}{n} \log\frac{\rhon(x)}{\sigman(x)} 
> a\Bigr\} \\
\quad& {\rm and } \quad
\Bigl\{x\in\cXn\;\Big|\; \frac{1}{n} \log\frac{\rhon(x)}{\sigman(x)} 
\geq a\Bigr\} .
\end{align*}
In general $S_n(a)$ may be randomized with an 
arbitrary probability 
when 
the obtained data $x$ satisfies 
$\frac{1}{n} \log\frac{\rhon(x)}{\sigman(x)} = a $. 
Denoting the characteristic functions 
of the sets $\{x\, |\, A(x) > c\}$ and $\{x\, |\, A(x) \geq c\}$ 
by $\{A>c\}$ and 
$\{A\geq c\}$ respectively, 
equation (\ref{classical_S_n(a)}) is rewritten as
\begin{equation}
\label{classical_S_n(a).var}
\left\{\frac{1}{n} \log\frac{\rhon}{\sigman} > a \right\}
\leq S_n(a) \leq
\left\{\frac{1}{n} \log\frac{\rhon}{\sigman} \geq a \right\} , 
\end{equation}
or equivalently as 
\begin{equation}
\label{Def_S_n(a).general}
\left\{\rhon - e^{na}\sigman > 0 \right\}
\leq S_n(a) \leq
\left\{\rhon - e^{na}\sigman \geq 0 \right\} .
\end{equation}
The family of tests $\{S_n (a)\}_{a\in\bR}$ 
characterizes the information spectrum by 
$ {\rm Prob} \{Z_n \ggeq a\} = \rhon [S_n(a)]$. 

In order to introduce the quantum analogue of $S_n (a)$, we 
need some preliminaries.  
For a  self-adjoint operator $A$ on a Hilbert space 
with the spectral 
decomposition\footnote{%
We assume here 
that $A$ has discrete eigenvalues since it suffices 
for our main concern and simplifies the description, 
although the assumption is not essential. 
}
 $A = \sum_{i} \lambda_i E_i$, where 
$\{\lambda_i\}$ are the eigenvalues and 
$\{E_i\}$ are the orthogonal 
projections onto the corresponding 
eigenspaces, we define 
\[ 
\{ A\geq 0\} \defeq \sum_{i:\lambda_i\geq 0} E_i 
\quad \mbox{and}\quad
\{ A > 0\} \defeq \sum_{i:\lambda_i > 0} E_i .
\]
These are the orthogonal projections onto 
the direct sum of eigenspaces corresponding to 
nonnegative and positive eigenvalues, respectively. 
The projections $\{ A\leq 0\}$ and $\{ A < 0\}$, 
or more generally $\{ A\geq B\} = \{A-B\geq 0\}$, 
$\{ A< B \}= \{A-B <0\}$, etc., 
are defined similarly. We have 
\begin{equation}
\label{TrA{A>0}>0}
\Tr (A \{ A > 0\}) \geq 0 , 
\end{equation}
and for any test $T$ on $\cH$ 
\begin{equation}
\label{Tr(A{A>0}geqTr (A T)}
\Tr (A \{ A > 0\}) \geq \Tr (A T) .
\end{equation}
The first inequality is obvious, while the second 
follows from $0\leq T\leq I$ as
\begin{eqnarray*}
\Tr (AT) &=& 
\Tr (A\{ A > 0\} T) + \Tr (A\{ A \leq 0\} T) \\
&\leq &\Tr (A\{ A > 0\} T) \\
&\leq& \Tr (A\{ A > 0\} ). 
\end{eqnarray*}
Note that $\{ A > 0\}$ in  (\ref{TrA{A>0}>0}) and 
(\ref{Tr(A{A>0}geqTr (A T)}) 
can be replaced with 
$\{ A \geq 0\}$ or, more generally, with any self-adjoint operator 
$S$ satisfying $\{ A > 0\}\leq S\leq \{ A \geq 0\}$.

Now, in the quantum setting where a sequence of 
density operators $\rhovec =\{\rhon\}$ and 
that of bounded nonnegative self-adjoint operators 
$\sigmavec =\{\sigman\}$ are given, 
let  $S_n(a)$ be a self-adjoint operator satisfying 
the same equation as (\ref{Def_S_n(a).general}). 
Since $S_n(a)$ satisfies $0\leq S_n(a)\leq I$, it is a test in our sense.  
Indeed, it is the quantum analogue of the likelihood test  
introduced by Holevo \cite{Ho72} and Helstrom \cite{Helstrom} 
when $\sigman$ is a density operator. 
Note that (\ref{Def_S_n(a).general}) is not equivalent to 
(\ref{classical_S_n(a).var}) in the quantum case 
unless $\rhon$ and $\sigman$ commute. 
As in the classical case, there is an ambiguity 
in the definition of $S_n(a)$, including two special cases 
$S_n(a) = \{\rhon - e^{na} \sigman \geq 0\}$ and 
$S_n(a) = \{\rhon - e^{na} \sigman > 0\}$.  
Some quantities defined in the sequel may depend on a choice 
of $S_n(a)$ within (\ref{Def_S_n(a).general}), but this  will not 
cause any essential difference in the theorems represented in terms 
of these quantities.  
We sometimes write $\{\rhon - e^{na} \sigman \ggeq 0\}$ 
to mean $S_n(a)$, 
suggesting this ambiguity. 
\par From (\ref{TrA{A>0}>0}) and (\ref{Tr(A{A>0}geqTr (A T)})  
we have 
\begin{equation}
\label{Positive__S_n(a)}
 (\rhon - e^{na}\sigman) \bigl[ S_n(a)\bigr] \geq 0, 
\end{equation}
and for any test $T_n$ 
\begin{equation}
\label{Optimal_S_n(a)}
(\rhon - e^{na}\sigman)  \bigl[ S_n(a)\bigr] \geq 
(\rhon - e^{na}\sigman) \bigl[ T_n \bigr]. 
\end{equation}
In addition, letting $S_n^{\,c}(a) \defeq I - S_n(a) = 
\{\rhon - e^{na} \sigman \lleq 0\}$ we have 
\begin{equation}
\label{Negative_S^c_n(a)}
 (\rhon - e^{na}\sigman) \bigl[ S^{\,c}_n(a)\bigr] \leq 0, 
\end{equation}
\begin{equation}
\label{Worst_S^c_n(a)}
(\rhon - e^{na}\sigman)  \bigl[ S^{\,c}_n(a)\bigr] \leq 
(\rhon - e^{na}\sigman) \bigl[ T_n^{\,c} \bigr] .
\end{equation}
These are rewritten as
\begin{equation}
\label{Positive__S_n(a).var}
\alphan (a) + e^{na} \betan(a) \leq 1, 
\end{equation}
\begin{equation}
\label{Optimal_S_n(a).var}
\alphan (a) + e^{na} \betan (a) \leq 
\alphan [T_n] + e^{na} \betan [T_n], 
\end{equation}
and
\begin{equation}
\label{Negative_S^c_n(a).var}
\alphan (a) - e^{na}\betancomp (a) \leq 0, 
\end{equation}
\begin{equation}
\label{Worst_S^c_n(a).var}
\alphan (a) - e^{na}\betancomp (a) \leq 
\alphan [\Tn] - e^{na}\betancomp [\Tn], 
\end{equation} 
where 
\begin{align*}
\alpha_n(a) &\defeq \alpha_n[S_n(a)] 
= \rho_n  \bigl[ \{\rho_n - e^{na}\sigman \lleq 0\}  \bigr], \\
\beta_n(a) &\defeq \beta_n[S_n(a)] 
= \sigma_n  \bigl[ \{\rho_n - e^{na}\sigman \ggeq 0\} \bigr], \\
\betancomp (a) &\defeq 
\betan[S_n^{\,c} (a)] 
= \sigma_n  \bigl[ \{\rho_n - e^{na}\sigman \lleq 0\} \bigr] .
\end{align*}
Needless to say, these properties also hold in the 
classical case.  In particular, 
the inequality (\ref{Optimal_S_n(a).var}) 
in the classical case is the Neyman-Pearson lemma, 
whose quantum extension was given in 
\cite{Ho72, Helstrom}. 
All the results in the later sections, including 
the classical ones obtained by Han, are derived 
only from the inequalities 
(\ref{Positive__S_n(a).var}) through (\ref{Worst_S^c_n(a).var}).  
This fact may be one of the most important findings 
of the present paper.  

Let us see that 
$\alpha_n(a)$ ($\beta_n(a)$, resp.) 
is monotonically nondecreasing (nonincreasing, resp.) 
as a function of $a$; i.e., if $a<b$ then 
\begin{equation}
\label{monotone_alpha_beta}
\alpha_n(a) \leq \alpha_n(b)
\quad \mbox{and}\quad
\beta_n(a) \geq \beta_n(b) .
\end{equation}
In the classical case, this is obvious because 
$\{x\, |\, \frac{1}{n} \log\frac{\rhon(x)}{\sigman(x)} 
> a\} \supset 
\{x\, |\, \frac{1}{n} \log\frac{\rhon(x)}{\sigman(x)} 
\geq b\}$ if $a<b$.  In order to show 
the monotonicity in the quantum case,  we invoke 
(\ref{Optimal_S_n(a).var}) to yield
\[
\alphan (a) + e^{na} \betan (a) \leq 
\alphan (b) + e^{na} \betan (b), \]
\[
\alphan (b) + e^{nb} \betan (b) \leq
\alphan (a) + e^{nb} \betan (a). 
\]
These are rewritten as 
\begin{equation}
\label{Appl_Optimal_S_n(a)}
e^{na} 
\left\{ \beta_n(a) - \beta_n(b) \right\} 
\leq 
\alpha_n(b) - \alpha_n(a) \leq 
e^{nb} 
\left\{ \beta_n(a) - \beta_n(b) \right\},  
\end{equation}
which leads to (\ref{monotone_alpha_beta}). 
This monotonicity will be used  implicitly 
throughout the later arguments. 

Let 
\begin{align*}
\etan(a) &\defeq 
\etan [S_n(a)] = - \frac{1}{n}\log\alpha_n(a) \\
&= -\frac{1}{n} \log \rhon[\{\rhon -e^{na}\sigman\lleq 0\}], \\
\zetan(a) 
&\defeq \zetan [S_n(a)] = - \frac{1}{n}\log\beta_n(a) \\
&= -\frac{1}{n} \log \sigman[\{\rhon -e^{na}\sigman\ggeq 0\}], \\
\zetancomp(a) 
&\defeq 
\zetancomp [S_n(a)] = - \frac{1}{n}\log \betancomp (a) \\
&= -\frac{1}{n} \log \sigman[\{\rhon -e^{na}\sigman\lleq 0\}],\\
\end{align*}
and 
\begin{align*}
\alphainf (a) &\defeq \alphainf [\Svec (a)] = 
\liminf_{n\rightarrow\infty}\alphan(a), \\
\alphasup (a) &\defeq \alphasup [\Svec (a)] = 
\limsup_{n\rightarrow\infty}\alphan(a), \\
\etainf(a)&\defeq \etainf[\Svec (a)] = 
\liminf_{n\rightarrow\infty}\etan(a), \\
\zetainf(a)&\defeq \zetainf[\Svec (a)] = 
\liminf_{n\rightarrow\infty}\zetan(a), \\
\zetasupcomp(a) &\defeq \zetasupcomp[\Svec (a)] = 
\limsup_{n\rightarrow\infty}\zetancomp(a) , 
\quad\text{etc.}, 
\end{align*}
where $\Svec (a)$ denotes the sequence $\{S_n(a)\}_{n=1}^\infty$. 
Note that $\etan(a), \zetancomp(a), \etainf (a)$ and $\zetasupcomp(a)$ 
are monotonically nonincreasing, while 
$\zetan(a)$ and $\zetainf(a)$ are monotonically 
nondecreasing. In addition, since (\ref{Positive__S_n(a).var})  yields 
$\betan(a)\leq e^{-na}$, we have
\begin{equation}
\label{zetainf_geq_a}
\zetainf (a) \geq a.
\end{equation}

\section{Asymptotics of Stein's type and 
spectral divergence rates}
\label{sec:stein}

In this section we treat  
the following quantities:
\begin{align*}
&B(\varepsilon\, |\,\rhovec \,\|\,\sigmavec) 
\defeq 
\sup_{\Tvec}\, \{\,\zetainf [\Tvec] \;|\; 
\alphasup[\Tvec] \leq\varepsilon \} \\
= &
\sup\, \{\, R\,|\, \exists\Tvec \in\vec{\cT} , \;  
\alphasup [\Tvec] \leq \varepsilon 
\sepand 
\zetainf [\Tvec] \geq R
\},  
\\
&B^\dagger(\varepsilon\, |\,\rhovec\,\|\,\sigmavec) 
\defeq 
\sup_{\Tvec}\, \{\,\zetainf [\Tvec] \,|\, 
\alphainf [\Tvec] < \varepsilon \}  \\
= &
\inf\, \{\, R\,|\, \forall \Tvec\in\vec{\cT}, \; 
{\rm if} \;\; 
\zetainf [\Tvec] \geq R \;\; {\rm then} \;\; 
\alphainf[\Tvec] \geq \varepsilon 
\}, 
\end{align*}
where $\varepsilon$ is a constant lying in the interval $[0, 1]$, and 
in particular 
\begin{align*}
&B(\rhovec\,\|\,\sigmavec) 
\defeq B(0\, |\,\rhovec\,\|\,\sigmavec) \\
=&
\sup_{\Tvec}\, \{\,\zetainf [\Tvec] \,|\, 
\lim_{n\rightarrow\infty} \alpha_n[T_n] = 0 \}
\\
=& \sup\, \{\, R\,|\, \exists\Tvec, \;  
\lim_{n\rightarrow\infty} \alpha_n[T_n] = 0 
\sepand 
\zetainf [\Tvec] \geq R
\} ,
\\
\label{Def_B+}
&B^\dagger (\rhovec\,\|\,\sigmavec) 
\defeq
B^\dagger (1\, |\,\rhovec\,\|\,\sigmavec) \\
= &
\sup_{\Tvec}\,  
\{\, \zetainf[\Tvec]\; |\; 
\alphainf [\Tvec] < 1
\} \\
= &
\inf\, \{\,R\,|\, \forall\Tvec, \; 
{\rm if} \;\;
\zetainf[\Tvec] \geq R 
\;\; {\rm then} \;\; 
\lim_{n\rightarrow\infty} \alpha_n [T_n] = 1
\} . 
\end{align*}
As will be seen in the next section, these quantities 
are the main concern of Stein's lemma in 
the classical i.i.d.\ case.  Note that we formally have 
$B(1 \, |\,\rhovec\,\|\,\sigmavec)=\infty$
and 
$B^\dagger(0 \, |\,\rhovec\,\|\,\sigmavec) = -\infty$, 
although they are of no importance. 
Obviously,  for any $0\leq\varepsilon_1 < \varepsilon_2\leq 1$
\[
B(\rhovec\,\|\,\sigmavec) \leq B(\varepsilon_1 \,
|\,\rhovec\,\|\,\sigmavec) \leq  B(\varepsilon_2 \,
|\,\rhovec\,\|\,\sigmavec) , 
\]
\[
B^\dagger(\varepsilon_1 \, |\,\rhovec\,\|\,\sigmavec) \leq 
B^\dagger(\varepsilon_2 \, |\,\rhovec\,\|\,\sigmavec) 
\leq 
B^\dagger (\rhovec\,\|\,\sigmavec), 
\]
and
\[
B(\rhovec\,\|\,\sigmavec) \leq 
B(\varepsilon_1 \, |\,\rhovec\,\|\,\sigmavec) \leq 
B^\dagger(\varepsilon_2 \, |\,\rhovec\,\|\,\sigmavec)
 \leq B^\dagger (\rhovec\,\|\,\sigmavec).
\]
In addition, $B(\varepsilon\, |\,\rhovec\,\|\,\sigmavec) $ 
is right continuous for any $0\leq\varepsilon <1$ 
in the sense that 
\begin{equation}
\label{B_rightcont}
B(\varepsilon\, |\,\rhovec\,\|\,\sigmavec) 
= 
\max_{\Tvec} \, \{\,\zetainf [\Tvec] \;|\; 
\alphasup[\Tvec] \leq\varepsilon \} 
= \inf_{\varepsilon'>\varepsilon} 
B(\varepsilon'\, |\,\rhovec\,\|\,\sigmavec) .
\end{equation}
To show this, let  $\{\delta_k\}_{k=1}^\infty$ be an arbitrary 
sequence of positive numbers satisfying $\lim_{k\rightarrow\infty} \delta_k 
= 0$.  
 Then for each $k$ 
there exist a test ${\Tvec}^{(k)} = \{T^{(k)}_n\}$ and 
  a number $n_k$ such that  
$\alphan [T^{(k)}_n ]\leq \varepsilon + 2 \delta_k$ and 
$\zetan [T^{(k)}_n] \geq 
B(\varepsilon+ \delta_k  \, |\,\rhovec\,\|\,\sigmavec) - \delta_k$  
for all $n\geq n_k$.  It is now easy to construct a test $\Tvec$ such that 
$\alphasup [\Tvec] \leq \varepsilon$ and $\zetainf [\Tvec] \geq 
\inf_{\varepsilon'>\varepsilon} 
B(\varepsilon'\, |\,\rhovec\,\|\,\sigmavec)$, which 
proves (\ref{B_rightcont}).  
On the other hand, it is obvious that 
$B^\dagger (\varepsilon\, |\,\rhovec\,\|\,\sigmavec)$ 
is left continuous 
for $0<\varepsilon \leq1$; 
\begin{equation}
\label{B_leftcont}
B^\dagger (\varepsilon\, |\,\rhovec\,\|\,\sigmavec) 
= \sup_{\varepsilon'<\varepsilon} 
B^\dagger (\varepsilon' \, |\,\rhovec\,\|\,\sigmavec). 
\end{equation}

Next, let 
\begin{align*}
\infD(\varepsilon\, |\, \rhovec\,\|\,\sigmavec) 
&\defeq 
\sup\, \{a\, |\, 
\alphasup (a) \leq \varepsilon\}, \\
\supD(\varepsilon\, |\, \rhovec\,\|\,\sigmavec) 
&\defeq
\sup\, \{a\, |\, 
\alphainf (a) < \varepsilon\} 
= \inf\, \{a\, |\, 
\alphainf (a) \geq \varepsilon\} 
\end{align*}
for $0\leq \varepsilon \leq 1$, and 
\begin{align*}
\infD (\rhovec\,\|\,\sigmavec) 
&\defeq \infD(0\, |\, \rhovec\,\|\,\sigmavec) 
= \sup\, \{\, a\,|\, 
\lim_{n\rightarrow\infty} \alphan (a) =0  \}, \\
\supD (\rhovec\,\|\,\sigmavec) 
&\defeq \supD(1\, |\, \rhovec\,\|\,\sigmavec) 
= \inf\, \{\, a\,|\, 
\lim_{n\rightarrow\infty} \alpha_n (a)  =1  \}.
\end{align*}
It should be noted that in the 
classical case when $\rhon = P_{\Xn}$ and $\sigman = P_{\altXn}$ 
we have 
\[
\infD(\rhovec\,\|\,\sigmavec)  = 
\mbox{p-}\liminf_{n\rightarrow\infty} 
\frac{1}{n} \log \frac{P_{\Xn}(\Xn)}{P_{\altXn}(\Xn)}
\]
and 
\[
\supD (\rhovec\,\|\,\sigmavec)  =
\mbox{p-}\limsup_{n\rightarrow\infty} 
\frac{1}{n} \log \frac{P_{\Xn}(\Xn)}{P_{\altXn}(\Xn)}, 
\]
where $\mbox{p-}\liminf$ and $\mbox{p-}\limsup$ are 
the liminf and limsup in probability: 
\begin{align*}
 \mbox{p-}\liminf_{n\rightarrow\infty} A_n 
&= \sup\, \{a\, |\, \lim_{n\rightarrow\infty }
{\rm Prob} \{ A_n \ggeq a\} = 1\}, \\
\mbox{p-}\limsup_{n\rightarrow\infty} A_n 
&= \inf\, \{a\, |\, \lim_{n\rightarrow\infty }
{\rm Prob} \{ A_n \ggeq a\} = 0\}.
\end{align*}
Actually,  
$\infD(\bX\,\|\,\altbX)$ and 
$\supD(\bX\,\|\,\altbX)$ 
were introduced in \cite{Han_book, Han_test} 
by these expressions and called the spectral sup- and inf-divergence rates 
between 
$\bX$ and $\altbX$.

It is clear that for any $0\leq\varepsilon_1 < \varepsilon_2\leq 1$ 
\[
\infD (\rhovec\,\|\,\sigmavec) \leq 
\infD (\varepsilon_1 \, |\,\rhovec\,\|\,\sigmavec) \leq 
\infD (\varepsilon_2 \, |\,\rhovec\,\|\,\sigmavec) , 
\]
\[
\supD (\varepsilon_1 \, |\,\rhovec\,\|\,\sigmavec) \leq 
\supD (\varepsilon_2 \, |\,\rhovec\,\|\,\sigmavec) 
\leq 
\supD  (\rhovec\,\|\,\sigmavec), 
\]
and
\[
\infD (\rhovec\,\|\,\sigmavec) \leq  
\infD (\varepsilon_1 \, |\,\rhovec\,\|\,\sigmavec) \leq 
\supD (\varepsilon_2 \, |\,\rhovec\,\|\,\sigmavec)
 \leq \supD (\rhovec\,\|\,\sigmavec).
\]
In addition, when $\sigmavec = \{\sigman\}$ consists of 
states (probability distributions in the classical case 
and density operators in the quantum case) 
as well as $\rhovec = \{\rhon\}$, 
 it follows from (\ref{Negative_S^c_n(a).var}) that 
\begin{equation}
\alphan (a) \leq e^{na}(1 -\betan (a) ) \leq e^{na}. 
\end{equation} 
Hence we have $\lim_{n\rightarrow\infty}\alphan (a) =0$ 
for any $a<0$, which leads to 
\begin{equation}
\infD (\rhovec\,\|\,\sigmavec )\geq 0. 
\end{equation}
On the other hand, $\infD (\rhovec\,\|\,\sigmavec )$ and 
$\supD (\rhovec\,\|\,\sigmavec )$ may be negative when 
$\{\sigma_n\}$ are not states.  

\begin{theorem}  \label{Th:SteinSpec}
For every $\varepsilon\in [0,1]$ 
\begin{eqnarray}
\label{B=infD_epsilon} 
B(\varepsilon\, |\,\rhovec\,\|\,\sigmavec) 
&=& 
\infD (\varepsilon\, |\,\rhovec\,\|\,\sigmavec), \\
\label{B^+=supD_epsilon} 
B^\dagger (\varepsilon\, |\,\rhovec\,\|\,\sigmavec) 
&=& 
\supD (\varepsilon\, |\,\rhovec\,\|\,\sigmavec) .
\end{eqnarray}
In particular, we have
\begin{eqnarray}
\label{B=infD} 
B(\rhovec\,\|\,\sigmavec)
 &=& 
\infD (\rhovec\,\|\,\sigmavec), \\
\label{B^+=supD} 
B^\dagger (\rhovec\,\|\,\sigmavec) 
&=& 
\supD (\rhovec\,\|\,\sigmavec) .
\end{eqnarray}
\end{theorem}

\begin{proof} 
Recalling  that $\alphan (a) = \alphan [S_n(a)]$ and 
using equation (\ref{zetainf_geq_a}), we have
\begin{eqnarray*}
B(\varepsilon\, |\, \rhovec\,\|\,\sigmavec) 
&\geq &
\sup_a\, \{\zetainf (a) \, |\, 
\alphasup (a) \leq\varepsilon\} \\
&\geq & 
\sup\, \{a \, |\, \alphasup (a) \leq\varepsilon\} 
= \infD (\varepsilon\, |\, \rhovec\,\|\,\sigmavec) .
\end{eqnarray*}
To show the converse inequality, 
suppose that a test $\Tvec$ and a
real number 
$a$ satisfy 
\[ \alphasup [\Tvec] 
\leq \varepsilon < 
\alphasup (a). \] 
Note that we can assume with no loss of generality 
the existence of such an $a$, or equivalently the finiteness of 
$\infD (\varepsilon\, |\, \rhovec\,\|\,\sigmavec)$, 
since the inequality is trivial otherwise. 
Then 
there exists a positive $\delta$ for which 
$ \alphan (a) - \alphan [T_n] \geq \delta$
holds for infinitely many $n$'s.  Using equation 
 (\ref{Optimal_S_n(a).var}) we have
\[ 
\betan [T_n] \geq e^{-na} 
\left\{\alphan (a) - \alphan [T_n]\right\} + \betan (a) 
\geq e^{-na} \delta
\]
for these $n$'s, which implies that 
$\zetainf [\Tvec] \leq a$.  This proves 
\[
B(\varepsilon\, |\, \rhovec\,\|\,\sigmavec) 
\leq 
\inf\, \{a\,|\, \alphasup (a) 
> \varepsilon \} 
= \infD (\varepsilon\, |\, \rhovec\,\|\,\sigmavec) .
\]
Equation (\ref{B=infD_epsilon}) has thus been verified. 
We can also prove equation (\ref{B^+=supD_epsilon}) 
almost in the same way.
\end{proof}

\begin{remark}
{\rm 
Equation (\ref{B=infD_epsilon}) for the classical case was obtained by  
Han \cite{Han_book} as a slight modification of a result by  Chen 
\cite{Chen}. 
Equation (\ref{B=infD}) was also described in \cite{Han_book}, giving 
credit to Verd\'{u} \cite{Verdu} for the original reference. 
As was mentioned in \cite{Han_book} and is now obvious from 
 (\ref{B=infD}) and (\ref{B^+=supD}), 
the equality $\infD (\rhovec\,\|\,\sigmavec) = 
\supD (\rhovec\,\|\,\sigmavec)$ 
is necessary and sufficient for the so-called strong converse 
property to hold 
in the sense that $\alphan[\Tn] $ converges to $1$ for 
any test $\Tvec$ satisfying 
$\zetainf[\Tvec] > B(\rhovec\,\|\,\sigmavec) $. 
}
\end{remark}

\section{Stein's lemma in the classical and quantum i.i.d.\ case}
\label{sec:i.i.d.}

In the classical i.i.d.\ case when 
$\rhon (x_1, \ldots , x_n) = \rho (x_1) \cdots \rho (x_n)$ 
and 
$\sigman (x_1, \ldots , x_n) = \sigma (x_1) \cdots \sigma (x_n)$, 
Stein's lemma (e.g., \cite{Blahut}, \cite{CovTho}, \cite{DemZei}) 
claims that 
\begin{equation}
\label{Stein}
B(\varepsilon_1\, |\, \rhovec\,\|\,\sigmavec) = 
B^\dagger (\varepsilon_2\, |\, \rhovec\,\|\,\sigmavec) 
= 
D(\rho\,\|\,\sigma) 
\end{equation}
for 
$0\leq \forall \varepsilon_1 < 1, \;\; 
0< \forall \varepsilon_2 \leq 1$,
where 
$D(\rho\,\|\,\sigma)$ is the Kullback-Leibler divergence: 
$D(\rho\,\|\,\sigma) = \sum_x \rho (x) \log\frac{\rho (x)}{\sigma (x)}$. 
A standard proof of the lemma uses a similar argument to the proof of 
Theorem~\ref{Th:SteinSpec} to reduce (\ref{Stein}) to 
\begin{equation}
\label{infD=supD=D}
\infD (\rhovec\,\|\,\sigmavec) = \supD(\rhovec\,\|\,\sigmavec) 
= D(\rho\,\|\,\sigma), 
\end{equation}
which is equivalent to 
\[ 
\lim_{n\rightarrow\infty}\frac{1}{n} 
\log \frac{\rhon (X_1, \ldots , X_n)}{\sigman (X_1,\ldots , X_n)} 
=
D(\rho\,\|\,\sigma)
\quad \mbox{in probability}, 
\]
where $X_1, \ldots , X_n$ are random variables obeying 
the probability distribution $\rhon$.  Now this  
is a direct consequence of 
the weak law of large numbers.

Let us turn to the quantum i.i.d.\ case when $\rhon = 
\rhotensor$ and $\sigman = \sigmatensor$, where 
$\rho$ and $\sigma$ are density operators on a Hilbert 
space $\cH$, and define  
the quantum relative entropy 
by $D(\rho\,\|\,\sigma) = \Tr [\rho (\log\rho -\log\sigma)]$ 
(e.g., \cite{OhyaPetz}). 
The achievability part 
\begin{equation}
\label{eq:direct_Stein}
B(\varepsilon\, |\,\rhovec\,\|\,\sigmavec ) 
\geq D(\rho\,\|\sigma) \quad\mbox{for}\quad 0<\forall \varepsilon <1,
\end{equation}
which is equivalent to 
$B(\rhovec\,\|\,\sigmavec ) \geq D(\rho\,\|\sigma)$ by 
(\ref{B_rightcont}), was first proved by Hiai and Petz \cite{Hiai-Petz}. 
They showed the existence of 
a sequence of POVMs $\Mvec=\{M^{(n)}\}$ on 
$\{\cHtensor\}$ satisfying 
\begin{equation}
\label{eq:HiaiPetz1}
  \liminf_{n\rightarrow\infty} \frac{1}{n} 
 D_{M^{(n)}}( \rhotensor\,\|\,\sigmatensor)
\geq  D(\rho\,\|\,\sigma),
\end{equation}
where $D_{M^{(n)}}(\rhotensor\,\|\,\sigmatensor)$ denotes 
the Kullback-Leibler divergence between the probability 
distributions 
$P_{\rhotensor}^{M^{(n)}} (\,\cdot\, ) = \Tr (\rhotensor M^{(n)}(\,\cdot\, 
))$ 
and 
$P_{\sigmatensor}^{M^{(n)}} (\,\cdot\, ) =\Tr (\sigmatensor 
M^{(n)}(\,\cdot\, ))$.  
Since $n D(\rho\,\|\,\sigma) = D( \rhotensor\,\|\,\sigmatensor)
\geq D_{M^{(n)}}( \rhotensor\,\|\,\sigmatensor)$ 
follows from the monotonicity of relative entropy, this leads to 
\begin{equation}
\label{eq:HiaiPetz}
 D(\rho\,\|\,\sigma) = \lim_{n\rightarrow\infty} \frac{1}{n} 
\sup_{M^{(n)}} D_{M^{(n)}}( \rhotensor\,\|\,\sigmatensor) , 
\end{equation}
which is often referred to as the Hiai-Petz theorem. 
Now it is easy to see that combination of  (\ref{eq:HiaiPetz1}) 
and the the direct part of the classical Stein's lemma 
leads to  (\ref{eq:direct_Stein}) as is shown in \cite{Hiai-Petz}. 
Hayashi \cite{Hayashi:Asympt} gave another construction of  
$\{M^{(n)}\}$ satisfying  (\ref{eq:HiaiPetz1}) 
based on a representation-theoretic consideration\footnote{%
More precisely, the papers  \cite{Hiai-Petz} and \cite{Hayashi:Asympt} 
showed different theorems, 
both of 
which include (\ref{eq:HiaiPetz1}) as a special case; 
see \cite{Hayashi:Asympt} for details. }. 
Inequality (\ref{eq:direct_Stein}) can also be 
proved more directly, not by way of (\ref{eq:HiaiPetz1}), 
in several different ways as shown in \cite{Hayashi:Optimal}, 
\cite{Ogawa-Hayashi} and Remark~20 of 
\cite{Hayashi-Nagaoka}, the last of which 
also appears in Sec.~3.6 of \cite{Hayashi:book}.  Note that these proofs 
conversely  
yield the existence of 
$\Mvec=\{M^{(n)}\}$ achieving (\ref{eq:HiaiPetz1}) with the help of 
(\ref{liminf_DTn}) below. 

On the other hand, 
the converse part 
$B^\dagger (\rhovec\,\|\,\sigmavec) \leq D(\rho\,\|\,\sigma)$ 
was first shown in \cite{Ogawa-Nagaoka} by 
combining the quantum Neyman-Pearson lemma (\ref{Optimal_S_n(a).var}) 
with the inequality\footnote{%
As a consequence of Eq.~(2.63) in \cite{Hayashi:book}, the inequality of 
(\ref{eq:OgaNag}) turns out to be true for $\forall \theta\geq 0$. 
} 
\begin{equation}
\label{eq:OgaNag}
\Tr \left(\rhotensor 
\{\rhotensor - e^{na}\sigmatensor \geq 0\} \right) 
\leq 
e^{- n \{as - \psi (\theta)\}} 
\end{equation}
for $0\leq\forall \theta\leq 1$,
where $\psi (\theta) \defeq \log\Tr (\rho^{1+\theta}\sigma^{-\theta})$. 
A simpler proof was given in \cite{Nagaoka:strong}. 

 The quantum Stein's lemma has thus been 
established in the same form as (\ref{Stein}).  
In the quantum case, (\ref{infD=supD=D}) is not a ground 
of  (\ref{Stein}), as at present we do not have a quantum version 
of the law of large numbers which directly applies to  (\ref{infD=supD=D}) 
(even though only the inequality 
$\supD(\rhovec\,\|\,\sigmavec) \leq D(\rho\,\|\,\sigma)$ 
follows immediately from (\ref{eq:OgaNag})). 
Instead, (\ref{infD=supD=D}) should be regarded as a consequence of  
(\ref{Stein}).  
So we restate it as a theorem. 

\begin{theorem}
\label{Thm:D=infD=supD}
For arbitrary density operators $\rho$ and $\sigma$ 
on a Hilbert space, we have 
$\infD (\rhovec\,\|\,\sigmavec) = \supD(\rhovec\,\|\,\sigmavec) 
= D(\rho\,\|\,\sigma)$ by letting $\rhovec=\{\rhotensor\}$ and 
$\sigmavec =\{\sigmatensor\}$; in other words, 
\begin{align}
\label{q_law_large_number}
&\lim_{n\rightarrow\infty} 
\Tr \left(\rhotensor \{\rhotensor - e^{na}\sigmatensor \ggeq 0\}\right) 
\nonumber\\
=& \left\{
\begin{array}{ccc}
1 &{\rm if}& a < D(\rho\,\|\sigma), \\
0 & {\rm if}& a > D(\rho\,\|\sigma) .
\end{array}
\right.
\end{align}
\end{theorem}

\begin{example} \ 
{\rm 
Let us numerically illustrate this theorem 
for 
\[
\rho = \left[ \begin{array}{cc}
0.75 & 0.35 \\
0.35 & 0.25
\end{array}\right] \sepand 
\sigma = \left[ \begin{array}{cc}
0.9 & 0 \\
0 & 0.1
\end{array}\right],
\]
which are density operators (matrices) on $\cH = \bC^2$.  
The relative entropy in this case is $D(\rho\|\sigma) = 0.4013\cdots$. 
The graph of the function 
$g_n(a)$ $\defeq$ 
$\Tr \left(\rhotensor \{\rhotensor - e^{na}\sigmatensor > 0\}\right)
$
for $n=5, 15$ and $50$ 
is shown in Fig.1, where we can see that 
the slope of the graph around $a= D(\rho\|\sigma)$ gets steeper  
with increase of $n$, 
as equation (\ref{q_law_large_number}) suggests. 
It is noted that drawing the graph requires computing 
the spectral decomposition of the $2^n\times 2^n$ matrix 
$\rhotensor - e^{na}\sigmatensor$ for each $a$, 
which is too large to apply a direct method when $n=50$. 
We have applied the theory of irreducible decomposition 
of the algebra generated by $\{A^{\otimes n}\;|\;A\in{\bC}^{2\times 2}\}$
based on the observation made in \cite{Hayashi:Asympt, Hayashi:Optimal}, 
which reduces the problem to finding the spectral decompositions of 
$\lceil(n+1)/2 \rceil$ matrices whose sizes are at most $(n+1)\times 
(n+1)$. 
Details of the algorithm will be reported elsewhere. 

\begin{figure}[htbp]
\begin{center}
\scalebox{0.8}{\includegraphics[scale=0.8]{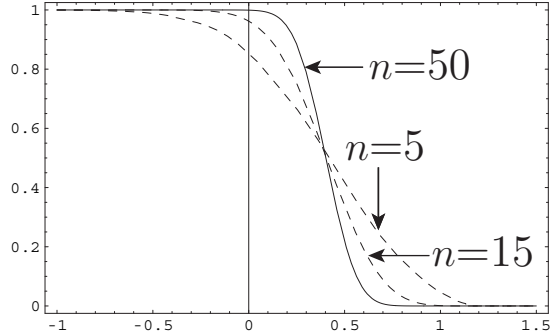}}
\end{center}
\caption{The graph of $g_n(a)$}
\label{graph1}
\end{figure}%
}
\end{example}

Let us make some observations on the quantum Neyman-Pearson test 
$S_n(a)$ in connection with (\ref{eq:HiaiPetz1}). 
 We begin by considering the general situation where $\rhovec=\{\rhon\}$ 
 and $\sigmavec=\{\sigman\}$ are arbitrarily given.  For a sequence 
of tests $\Tvec =\{\Tn\}$, 
let $D_{\Tn} (\rhon\,\|\,\sigman)$ denote 
the Kullback-Leibler divergence 
of the resulting probability distributions 
$(\rhon [\Tn], 1-\rhon[\Tn])$ and $(\sigman [\Tn], 1-\sigman[\Tn])$.  
Then we have
\begin{align*}
&D_{\Tn} (\rhon\,\|\,\sigman) \\
=&
- h(\rhon[\Tn]) -\rhon[\Tn] \log\sigman [\Tn] \\
&\quad - (1-\rhon[\Tn]) \log (1-\sigman [\Tn] ) \\
\geq &
-\log 2 - \rhon[\Tn] \log\sigman [\Tn] , 
\end{align*}
where $h$ is the binary entropy function; 
$h (t) \defeq -t\log t - (1-t) \log (1-t)$. 
This proves that 
\begin{equation}
\liminf_{n\rightarrow\infty}\frac{1}{n} D_{\Tn} (\rhon\,\|\,\sigman) 
\geq \zetainf [\Tvec]
\quad\mbox{if}\quad 
\lim_{n\rightarrow\infty} \alphan[\Tn] =0.
\label{liminf_DTn}
\end{equation}
In particular, we have 
\begin{equation}
\liminf_{n\rightarrow\infty}\frac{1}{n} D_{S_n(a)} (\rhon\,\|\,\sigman) 
\geq a 
\quad\mbox{if}\quad 
a < \infD(\rhovec\,\|\,\sigmavec), 
\label{liminf_DSn}
\end{equation}
where we have used (\ref{zetainf_geq_a}). 
Applying this to 
the i.i.d.\ case where $\rhon = \rho^{\otimes n}$ 
and $\sigman = \sigma^{\otimes n}$, and 
recalling Theorem~\ref{Thm:D=infD=supD}, we have 
\[
\liminf_{n\rightarrow\infty}\frac{1}{n} D_{S_n(a)} (\rhotensor 
\,\|\,\sigmatensor) 
\geq a 
\quad\mbox{if}\quad 
a < D(\rho\,\|\,\sigma) .
\]
Therefore, if a number sequence 
$\{a_n\}$ is chosen so that $a_n$ 
converges to $D(\rho\,\|\,\sigma)$ 
monotonically from below with 
a sufficient slow speed, then 
\begin{equation}
\liminf_{n\rightarrow\infty} 
\frac{1}{n}
D_{S_n(a_n)} (\rho^{\otimes n}\,\|\,\sigma^{\otimes n}) 
\geq D(\rho\,\|\,\sigma),
\end{equation}
which gives an example of (\ref{eq:HiaiPetz1}). 

\begin{remark}
{\rm 
From (\ref{liminf_DSn}) and the monotonicity of quantum relative entropy, 
we obtain the general inequality
\[
\liminf_{n\rightarrow\infty}\frac{1}{n} D (\rhon\,\|\,\sigman) \geq 
\infD(\rhovec\,\|\,\sigmavec), 
\]
which has also appeared in \cite{Hayashi-Nagaoka}. 
}
\end{remark}

\begin{remark}
\label{remark:sanov}
{\rm 
Recently an extension of quantum Stein's lemma was reported under the name 
of ``a quantum version of Sanov's theorem" \cite{BjeEtAl}.  Relating to 
this work, 
let us make some remarks on the relation between Stein's lemma and Sanov's 
theorem
(see e.g.\ \cite{DemZei}). 
These theorems are similar in that both of them represent the 
convergence rates of some probabilities in terms of relative entropy. 
Moreover, they are closely related to each other in their logical 
derivations.  Nevertheless, 
we should note that they have their respective 
roles in different contexts in general; 
Stein's lemma is about hypothesis testing and Sanov's theorem 
is a fundamental theorem in large deviation theory for empirical distribution. 
We also note that 
distinction of their roles is indispensable for precise 
understanding of both the significance of the Neyman-Pearson lemma and 
that of empirical distributions. Even though the result of \cite{BjeEtAl} 
has a certain significance from a viewpoint of 
hypothesis testing, its formulation does not precisely correspond to that 
of Sanov's 
theorem in the classical case. 
Finding a meaningful and useful quantum extension of Sanov's theorem is 
still a challenging 
open problem.  See also Remark~\ref{remark:Be_quant} below. 
}
\end{remark}

\section{Tradeoff between the exponents 
of the first and second kind error probabilities}
\label{sec:error_exp}

In order to properly evaluate the tradeoff between 
the error exponents of the first and second kinds 
for the classical hypothesis testing problem, 
Han \cite{Han_book, Han_test} introduced the following quantity:
\begin{align}
&B_e(r\,|\, \rhovec\,\|\,\sigmavec) \nonumber\\
\defeq& 
\sup\, \{\,R\,|\, \exists\Tvec\in\vec{\cT}, \;
\etainf[\Tvec] \geq r 
\sepand 
\zetainf[\Tvec] \geq R
\}\label{Def_Be} \\
\nonumber
=&
\sup_{\Tvec}\, \{\,\zetainf[\Tvec]\,|\,
\etainf [\Tvec]\geq r\}, 
\end{align}
where the subscript $e$ is intended to 
mean that the quantity concerns exponents. 
Roughly speaking, the second kind error probability 
optimally tends to $0$ with  the rate 
$\betan [\Tn] 
\approx e^{- n B_e(r\,|\, \rhovec\,\|\,\sigmavec)}$
when the first error probability is required to tend 
to $0$ with  $\alphan [\Tn] \approx 
e^{-n r}$ or faster. The same definition is applied to 
our setting including generalized and quantum hypothesis 
testing problem.  We shall give some characterizations 
to this quantity in the sequel, extending the formula obtained by Han. 
The following lemma will play an essential role. 

\begin{lemma}
\label{lemma:eta_and_zeta}
For any real number $a$ and any 
sequence of tests $\Tvec$ we have 
\begin{align}
\label{eq:eta_and_zeta.1}
\zetainf (a) \geq& 
\min\,\{ \zetainf[\Tvec] , \, a + \etainf [\Tvec] \,\}, 
\quad \mbox{and} \\
\label{eq:eta_and_zeta.2}
\etainf (a) \geq& 
\min\,\{ \etainf  [\Tvec] , \, 
- a+ \zetainf [\Tvec] \} . 
\end{align}
In particular, for any $a < b$
\begin{align}
\label{eq:eta_and_zeta.3}
\zetainf (a) \geq a + \etainf (b) 
\quad &\mbox{if} \quad \zetainf (a) < \zetainf (b), 
\quad \mbox{and} \\
\label{eq:eta_and_zeta.4}
\etainf (b) \geq 
-b + \zetainf (a) 
\quad &\mbox{if}\quad \etainf (a) > \etainf (b) . 
\end{align}
\end{lemma}

\begin{proof}
We have 
\begin{align*}
&\zetainf (a) \\
=&
- \limsup_{n\rightarrow\infty}
\frac{1}{n} \log \betan (a) \\
\stackrel{({\rm i})}{=}&
- \max \left\{
\limsup_{n\rightarrow\infty}\frac{1}{n} \log
\betan [\Tn], 
\;
\limsup_{n\rightarrow\infty} \frac{1}{n} \log
\Bigl(
\betan(a) - \betan[\Tn] 
\Bigr)\right\} \\
=&
\min\left\{
\zetainf [\Tvec], \; 
\liminf_{n\rightarrow\infty} -\frac{1}{n}\log 
\Bigl[
\betan(a) - \betan[\Tn]
\Bigr]\right\} \\
\stackrel{({\rm ii})}{\geq}&
\min\,\{ \zetainf[\Tvec] , \, a + \etainf [\Tvec] \,\}, 
\end{align*}
where the equality $\stackrel{({\rm i})}{=}$ follows from the formula
\begin{align}
&\limsup_{n\rightarrow\infty} \frac{1}{n} \log 
(x_n + y_n) \nonumber\\
= &
\max\left\{ 
\limsup_{n\rightarrow\infty} \frac{1}{n} \log x_n , 
\;
\limsup_{n\rightarrow\infty} \frac{1}{n} \log y_n 
\right\} 
\label{formula_limsup}
\end{align}
which is valid for any sequences of positive numbers 
$\{x_n\}, \{y_n\}$,  and 
the inequality $\stackrel{({\rm ii})}{\geq}$
follows from (\ref{Optimal_S_n(a).var}). 
The inequality (\ref{eq:eta_and_zeta.1}) is 
thus proved.  The proof of  (\ref{eq:eta_and_zeta.2}) is 
similar and omitted. 
\end{proof}

\begin{theorem} 
\label{thm:Be}
For any $r\geq 0$, 
\begin{align*}
B_e(r\,|\, \rhovec\,\|\,\sigmavec) 
&= 
\sup_a\, 
\{\,\zetainf (a)\;|\; \etainf (a) \geq r\,\} \\
&=
\inf_a\, \{\, a + \etainf (a)\;|\; \etainf (a) < r\,\}  \\
&= 
\zetainf (a_0 - 0) 
=
a_0 + \etainf (a_0 + 0) ,
\end{align*}
where 
\begin{align*}
& a_0 (\in\bR\cup \{-\infty, \infty\}) \\
\defeq &\sup\, \{a'\,|\, \exists a,\; \etainf (a)\geq r 
\;\;{\rm and}\;\; \zetainf (a') = \zetainf (a)  \,\}.
\end{align*}
\end{theorem}

\begin{remark}
{\rm 
Throughout the paper we use the notation 
\[ f(a+0) = \lim_{\varepsilon\downarrow 0} f(a+\varepsilon), 
\quad 
f(a-0) = \lim_{\varepsilon\downarrow 0} f(a-\varepsilon), 
\]
\[
f(\infty) = \lim_{a\rightarrow\infty} f(a), 
\quad 
\quad 
f(-\infty) = \lim_{a\rightarrow -\infty} f(a) 
\]
for a monotone function 
$f :\bR\rightarrow \bR\cup \{-\infty, \infty\}$. 
}
\end{remark}

\begin{remark}
{\rm 
The formula given in the above theorem 
is valid even in the `singular' 
case when the set 
$\{ a\,|\, \etainf (a) \geq r\}$ is 
empty or the entire real line $\bR$.  When the set 
is empty, we have $B_e(r\,|\, \rhovec\,\|\,\sigmavec) =
- \infty$, although this does not occur in the case when 
$\sigman$ are states (i.e., \ $\sigman [I] = 1$).  
When the set is $\bR$, we have 
$B_e(r\,|\, \rhovec\,\|\,\sigmavec) =
\infty$.}
\end{remark}

\noindent{\em Proof of Theorem \ref{thm:Be}:} 
\quad 
Since  $\etainf(a) = \etainf[\Svec(a)]$ 
and $\zetainf(a) = \zetainf[\Svec(a)]$, 
it is immediate from the definition 
(\ref{Def_Be}) of 
$B_e(r\,|\, \rhovec\,\|\,\sigmavec)$ 
that 
\begin{equation}
\label{eq:proof_Be.1}
B_e(r\,|\, \rhovec\,\|\,\sigmavec)
\geq
\sup_a\, \{\,\zetainf (a)\;|\; \etainf (a) \geq r\,\} . 
\end{equation}
Also, we immediately have
\begin{align}
\nonumber
&\sup_a\, \{\,\zetainf (a)\;|\; \etainf (a) \geq r\,\}  \\
\nonumber
&=
\label{eq:proof_Be.2}
\sup_{a'}\, \{\,\zetainf (a')\;|\; 
\exists a,\; \etainf (a)\geq r 
\;\;{\rm and}\;\; \zetainf (a') = \zetainf (a) \,\} \\
&\geq \zetainf (a_0 - 0) 
\end{align}
and 
\begin{equation}
\label{eq:proof_Be.3}
a_0 + \etainf (a_0 + 0) 
\geq 
\inf_a\, \{\, a + \etainf (a)\;|\; \etainf (a) < r\,\} .
\end{equation}
Next, 
for any sequence of 
tests $\Tvec = \{T_n\}$ satisfying 
$
\etainf[\Tvec] \geq r 
$
and for any number $a$ satisfying 
$\etainf(a) < r$, it follows from (\ref{eq:eta_and_zeta.2}) that
$
\zetainf[\Tvec] \leq a + \etainf(a)
$, 
which proves 
\begin{equation}
\label{eq:proof_Be.4}
B_e(r\,|\, \rhovec\,\|\,\sigmavec) 
\leq 
\inf_a\, \{\, a + \etainf (a)\;|\; \etainf (a) < r\,\} .
\end{equation}
Finally, let us show that 
\begin{equation}
\label{eq:proof_Be.5}
\zetainf (a_0-0) \geq  a_0 + \etainf (a_0+0) ,
\end{equation}
which, combined with (\ref{eq:proof_Be.1})
through (\ref{eq:proof_Be.4}), 
completes the proof of the theorem.
Assume first that the set 
$\{a\,|\,\etainf (a) \geq r\}$ is not empty 
nor $\bR$.  Then, letting $b$ be an 
arbitrary number satisfying $\etainf (b) < r$, 
it follows  from (\ref{eq:eta_and_zeta.4}) that 
\[ 
\sup_a\, 
\{\,\zetainf (a)\;|\; \etainf (a) \geq r\,\} 
\leq b + \etainf (b) < b + r < \infty. 
\]
Therefore, invoking that 
$\lim_{a\rightarrow\infty} \zetainf (a) = 
\infty$ follows from (\ref{zetainf_geq_a}), 
we see that $a_0$ is not $\infty$ nor $-\infty$. 
Now, for an arbitrary $\varepsilon >0$,  we have 
$\zetainf (a_0-\varepsilon) < \zetainf (a_0+\varepsilon) $
by the definition of $a_0$.  Hence,  from 
(\ref{eq:eta_and_zeta.3}) of 
Lemma~\ref{lemma:eta_and_zeta} we have
\[
\zetainf (a_0-\varepsilon) \geq 
a_0-\varepsilon + \etainf (a_0+\varepsilon), 
\] 
which leads to (\ref{eq:proof_Be.5}). 
When $\{a\,|\,\etainf (a) \geq r\} = \phi$, 
on the other hand, we have $a_0 = -\infty$, and 
(\ref{eq:proof_Be.5}) is obvious since 
the right-hand side is $-\infty$.  
When 
 $\{a\,|\,\etainf (a) \geq r\} = \bR$, 
we have $a_0= \infty$ and, again,  (\ref{eq:proof_Be.5}) is obvious since 
the left-hand side is $\infty$. 
 \hfill\QED\bigskip

\begin{remark}
{\rm 
The formula 
$B_e(r\,|\, \rhovec\,\|\,\sigmavec) 
= 
\inf_a\, \{\, a + \etainf (a)\;|\; \etainf (a) < r\,\} $
for the classical hypothesis testing was derived by 
Han \cite{Han_book, Han_test}, 
using a more complicated argument 
based on the technique of information-spectrum slicing. 
Note that this technique consists of a procedure of partitioning 
a set and does not straightforwardly apply 
to the quantum setting.  
Theorem~3 of \cite{Chen} was also intended to 
give a general formula for the tradeoff 
between the error exponents 
of the first and second kinds, 
but its proof contains 
a gap, and the theorem 
does not apply to the general 
case where $\etainf (a)$ and $\zetainf (a)$ 
may be discontinuous functions; see 
Example 3.6 of \cite{Han_test}.
}
\end{remark}

\begin{example}
{\rm 
Let us consider the case when $\rhovec$ and $\sigmavec$ 
consist of 
pure quantum sates of the form $\rhon = |\psin\rangle \langle\psin |$ 
and $\sigman = |\varphin\rangle \langle\varphin |$ , where 
$\psin$ and $\varphin$ are unit vectors in a Hilbert space 
$\cHn$ for every $n$, and let $\deltan \defeq \Tr (\rhon \sigman) 
= |\langle\psin | \varphin\rangle |^2$.  
In order to treat the hypothesis testing problem for this 
situation, 
we can assume 
with no loss of generality that 
\[ 
\psin = \left[
\begin{array}{c} 1 \\ 0 \end{array}
\right] 
\quad\mbox{and}\quad 
\varphin = \left[
\begin{array}{c} \sqrt{\deltan} \\ \sqrt{1-\deltan} \end{array}
\right] ,
\]
for which we have the following 
spectral decomposition:
\[ 
\rhon - e^{na}\sigman = 
\lambda_1 E_1 + \lambda_2 E_2, 
\]
where
\[
\lambda_1 = \frac{1-e^{na}}{2} + r, 
\quad 
\lambda_2 = \frac{1-e^{na}}{2} - r, 
\]
\[
E_1 = \frac{1}{2} 
\left[ \begin{array}{cc}
1+ \frac{1+e^{na}-2e^{na}\deltan}{2r} & 
- \frac{e^{na}\sqrt{\deltan (1-\deltan)}}{r} \\
& \\
- \frac{e^{na}\sqrt{\deltan (1-\deltan)}}{r} & 
1- \frac{1+e^{na}-2e^{na}\deltan}{2r} 
\end{array}
\right], 
\]
\[
E_2 = \frac{1}{2} 
\left[ \begin{array}{cc}
1- \frac{1+e^{na}-2e^{na}\deltan}{2r} & 
 \frac{e^{na}\sqrt{\deltan (1-\deltan)}}{r} \\
& \\
 \frac{e^{na}\sqrt{\deltan (1-\deltan)}}{r} & 
1+ \frac{1+e^{na}-2e^{na}\deltan}{2r} 
\end{array}
\right] , 
\]
with 
\[ r \defeq \frac{\sqrt{(1+e^{na})^2 - 4 e^{na} \deltan}}{2} .
\]
This leads to $\{ \rhon - e^{na}\sigman >0\} = E_1$ and 
\[ 
\Tr \left( \rhon \{ \rhon - e^{na}\sigman >0 \} \right) 
= 
\frac{1}{2} + 
\frac{1+  e^{na}-2  e^{na} \deltan}
{2\sqrt{(1+e^{na})^2 - 4  e^{na}\deltan}} .
\]
Thus we have for every $a>0$
\begin{align*}
\lim_{n\rightarrow\infty}
\Tr \left( \rhon \{ \rhon - e^{na}\sigman >0 \} \right) 
=1 
\quad \Longleftrightarrow \quad
\lim_{n\rightarrow\infty}\deltan =0 , 
\\
\lim_{n\rightarrow\infty}
\Tr \left( \rhon \{ \rhon - e^{na}\sigman >0 \} \right) 
=0 
\quad \Longleftrightarrow \quad
\lim_{n\rightarrow\infty}\deltan =1 , 
\end{align*}
which yields
\[
\infD (\rhovec\,\|\,\sigmavec) = 
\left\{ 
\begin{array}{cl}
\infty & 
\displaystyle
\mbox{if}\;\; \lim_{n\rightarrow\infty} \deltan =0, \\
0 & \mbox{otherwise}, 
\end{array} \right.  
\]
\[
\supD (\rhovec\,\|\,\sigmavec) = 
\left\{ 
\begin{array}{cl}
0 & 
\displaystyle
\mbox{if}\;\; \lim_{n\rightarrow\infty} \deltan =1, \\
\infty & \mbox{otherwise}.
\end{array} \right.
\]
Furthermore, it is not difficult to see that for every $a>0$ 
\[
\etainf (a) = -\limsup_{n\rightarrow\infty} \frac{1}{n}\log\deltan ,
\]
and letting $c$ denote this constant $\etainf (a)$, we have
\begin{equation}
\label{Be_pure} 
B_e (r\,|\,\rhovec\,\|\,\sigmavec ) = 
\left\{ 
\begin{array}{cl}
\infty &\mbox{if}\;\; r\leq c , \\
c & \mbox{if}\;\; r >c .
\end{array} \right. 
\end{equation}
Actually, the test $T^{(1)}_n \defeq\rhon = |\psin\rangle \langle\psin |$ 
satisfies $\alphan [T^{(1)}_n] = 0$ and $\betan[T^{(1)}_n] = \deltan$ for 
all $n$ and hence $\etainf[{\Tvec}^{(1)}] = \infty$  and $\zetainf 
[{\Tvec}^{(1)}] =c$, 
while 
the test $T^{(2)}_n \defeq \sigman = |\varphin\rangle \langle\varphin |$ 
satisfies 
$\etainf[{\Tvec}^{(2)}] = c$ and $\zetainf [{\Tvec}^{(2)}] =\infty$ . 
Equation (\ref{Be_pure}) means that it suffices to consider only 
these extreme tests when our concern is limited to 
the exponents of the error probabilities.  
In the i.i.d.\ case where $\cHn = \cH^{\otimes n}, 
\psin = \psi^{\otimes n}$ and $\varphin = \varphi^{\otimes n}$ 
for distinct unit vectors $\psi$ and $\varphi$, 
we have $\infD (\rhovec\,\|\,\sigmavec) = \supD(\rhovec\,\|\,\sigmavec) = 
\infty$, which  
is also seen from $D(\rho\,\|\,\sigma) = \infty$ together with   
the argument of section \ref{sec:i.i.d.}, and $c= -\log |\langle\psi 
|\varphi\rangle |^2$. 
}
\end{example}

\begin{remark}
\label{remark:Be_quant}
{\rm 
In the classical i.i.d.\ case, it follows from 
Sanov's theorem and Cram\'{e}r's theorem in large deviation theory 
that, for $- D(\sigma\,\|\, \rho) \leq \forall a \leq D(\rho\,\|\,\sigma)$, 
\[ \etainf (a) = \etasup (a) 
= 
\min_{\tau\,:\,\tau[\log\rho - \log\sigma] \leq a}
D(\tau\,\|\,\rho)
=
\max_{\theta\in\bR} (\theta a -\psi (\theta)), 
\]
where $\psi(\theta) \defeq \log \sum_x 
\rho(x)^{1+\theta}\sigma(x)^{-\theta}$, 
and 
\[ \zetainf (a) = \zetasup (a) 
 = 
\min_{\tau\,:\,\tau[\log\rho - \log\sigma] \geq a} 
D(\tau\,\|\,\sigma)
 = a + \overline{\underline{\eta}} (a) .
\]
Applying these relations to Theorem~\ref{thm:Be} 
with some additional calculations, we can 
derive several 
single-letterized expressions\footnote{%
The expressions (the first one in (\ref{eq:hoeffding}) in particular) 
are often referred to as {\em Hoeffding's theorem} after \cite{Hoe}. 
}
for $B_e (r\,|\,\rhovec\,\|\,\sigmavec )$ 
(see \cite{Han_book}, \cite{Han_test}, \cite{Blahut}, \cite{DemZei}), 
among which are
\begin{align}
B_e (r\,|\,\rhovec\,\|\,\sigmavec )
=& \min_{\tau\,:\,D(\tau\,\|\,\rho) \leq r} D(\tau\,\|\,\sigma)\nonumber\\
= &\max_{-1\leq\theta <0}\frac{(1+\theta)r +\psi(\theta)}{\theta} .
\label{eq:hoeffding}
\end{align}
In the quantum i.i.d.\ case, on the other hand,
we have no explicit formulas  
for $\overline{\underline{\eta}} (a)$, $\overline{\underline{\zeta}} (a)$ 
and $B_e (r\,|\,\rhovec\,\|\,\sigmavec )$ at present; see 
\cite{Ogawa-Hayashi} and \cite{Hayashi:book} (section~3.4) for 
some partial results\footnote{%
Note that $B(r\,|\rho\,\|\,\sigma)$ in \cite{Hayashi:book} 
corresponds to our $B_e (r\,|\,\sigmavec\,\|\,\rhovec )$. 
}.  
The mathematical difficulty arising in the study of $B_e 
(r\,|\,\rhovec\,\|\,\sigmavec )$ 
is closely related to the absence of a ``quantum large deviation theorem" 
applicable to $\overline{\underline{\eta}} (a)$ and 
$\overline{\underline{\zeta}} (a)$ 
(cf.\ Remark~\ref{remark:sanov}). 
}
\end{remark}

\section{Exponents of probability of correct testing}
\label{sec:correct_exp}

Suppose that $\{\sigman\}$ are states 
(i.e., $\sigman [I] =1$) and let $r$ be 
a real number greater than $B^\dagger (\sigmavec\,\|\,\rhovec) 
= \supD(\sigmavec\,\|\,\rhovec)$. 
When the first kind error probability $\alphan[\Tn]$ 
 of a sequence of tests $\Tvec$ 
tends to $0$ with a speed not slower than 
$e^{-nr}$, 
the second kind error probability $\betan[\Tn]$ inevitably 
tends to $1$. 
In this case, the speed  at which the probability of 
correct testing $1-\betan[\Tn]$ tends to $0$ 
can be regarded as a measure to evaluate ``badness" of $\{\Tn\}$. 
Hence, it is meaningful to investigate the slowest 
convergence rate of $1-\betan[\Tn]$ 
when $\alphan[\Tn]$ is required to tend to $0$ 
with $\alphan[\Tn]\approx e^{-nr}$ or faster. 
We are thus led to introduce the 
following quantity:
\begin{equation}
\label{Def_Be*}
B_e^*(r\,|\, \rhovec\,\|\,\sigmavec) 
\defeq 
\inf_{\Tvec}\, \{\,\zetasupcomp [\Tvec] 
\;|\; \etainf[\Tvec] \geq r\} ,
\end{equation}
where 
\[ 
\zetasupcomp [\Tvec] = 
\limsup_{n\rightarrow\infty} \;
\{-\frac{1}{n}\log (1-\betan[\Tn])\}. 
\]
Note that 
$B_e^*(r\,|\, \rhovec\,\|\,\sigmavec) $ 
is defined for every $r$ but is meaningless 
when $r< B^\dagger (\sigmavec\,\|\,\rhovec) 
= \supD(\sigmavec\,\|\,\rhovec)$ 
since it vanishes for such an $r$. 
Han \cite{Han_book, Han_test} introduced $B_e^*(r\,|\, \rhovec\,\|\,
\sigmavec) $  
for the classical hypothesis testing problem 
and characterized it as
 \begin{equation}
\label{Han_B^*}
B_{e}^*(r\,|\, \rhovec\,\|\,\sigmavec) 
=
\inf_a \Bigl\{ a+ \eta (a) + 
[r-\eta (a)]_+\Bigr\}, 
\end{equation}
where $[\, t\, ]_+  \defeq \max\,\{t, 0\}$, 
assuming the two conditions that the limit 
$\eta (a)  \defeq 
\lim_{n\rightarrow\infty} 
\etan (a) $ exists for all $a$ 
and that for any $M$ there exists a $K$ such that 
\[
\liminf_{n\rightarrow\infty} 
- \frac{1}{n} \log 
\sigman 
\Bigl[\left\{\frac{1}{n} \log \frac{\sigman}{\rhon}\geq K\right\}\Bigr] 
\geq M, 
\]
or equivalently
\begin{equation}
\label{unbounded_zetainfcomp}
\zetainfcomp (-\infty) = \infty.
\end{equation}

In this section we provide $B_{e}^*(r\,|\, \rhovec\,\|\,\sigmavec)$ 
with  a new characterization which needs no extra condition.  
Having in mind both applicability to source coding problems and 
consistency with the notation in \cite{Han_book, Han_test}, 
we exchange the roles of $\rhovec$ and $\sigmavec$ assumed 
in (\ref{cond:rhon}) and (\ref{cond:sigman}), so that 
\begin{equation}
0 < \rhon[I] \leq\infty
\quad\mbox{and}\quad
\sigman[I] = 1
\end{equation}
are now assumed.  Accordingly,  the definition of $\alphan[\Tn]$ in 
(\ref{alphanTn}) 
is changed 
into $\alphan[\Tn]\defeq \rhon[I-T_n]$ together with those of $\etan[\Tn], 
\etan (a), 
\etainf[\Tvec]$ and $\etainf (a)$.  The arguments below 
are based on the inequalities 
(\ref{Negative_S^c_n(a).var}) and (\ref{Worst_S^c_n(a).var}), 
which do not suffer from this change. 

\begin{lemma}
\label{lemma:eta_and_zeta^c}
For any real number $a$ and any 
sequence of tests $\Tvec$ we have 
\begin{equation}
\label{eq:eta_and_zeta^c.1}
\zetasupcomp[\Tvec] 
\geq
\min \,\{ \zetasupcomp (a), \; a+ \etainf[\Tvec] \,\} .
\end{equation}
In particular, for any $a < b$
\begin{equation}
\label{eq:eta_and_zeta^c.2}
\zetasupcomp (b) \geq a +  \etainf (b) 
\quad\mbox{if}\quad 
\zetasupcomp (a) > \zetasupcomp (b).
\end{equation}
\end{lemma}

\begin{proof}
We have 
\begin{align*}
\zetasupcomp[\Tvec] 
&= \limsup_{n\rightarrow\infty} 
- \frac{1}{n} \log \betancomp [\Tn] \\
&\geq
\limsup_{n\rightarrow\infty}
- \frac{1}{n} \log \Bigl( 
\betancomp (a) + e^{-na} \alphan [T_n]\Bigr) \\
&\geq
\min \left\{ \zetasupcomp (a), \; a + \etainf[\Tvec]  \right\} , 
\end{align*}
where the first inequality follows from 
 (\ref{Worst_S^c_n(a).var}) and the 
second inequality follows from 
the general formula 
(cf.\ (\ref{formula_limsup})):
\begin{align}
&\liminf_{n\rightarrow\infty} \frac{1}{n} \log 
(x_n + y_n) \nonumber \\
\leq  &
\max\left\{ 
\liminf_{n\rightarrow\infty} \frac{1}{n} \log x_n , 
\;
\limsup_{n\rightarrow\infty} \frac{1}{n} \log y_n 
\right\} . 
\label{formula_liminf}
\end{align}
\end{proof}

\begin{theorem} 
\label{thm:Be*}
For any real number $r$, we have
\begin{align}
B_e^*(r\,|\, \rhovec\,\|\,\sigmavec)
&=  B_{e,1}^*(r\,|\, \rhovec\,\|\,\sigmavec) \\
&= \sup_a \; \min\,\{\zetasupcomp (a), \, r+a\} \\
&= \inf_a \; \max\, \{\zetasupcomp (a), \, r+a\} \\
&= r + a^*_0, 
\end{align}
where
\begin{equation}
B_{e,1}^*(r\,|\, \rhovec\,\|\,\sigmavec) 
\defeq 
\inf_{\Tvec}\, \{\,\zetasupcomp [\Tvec] 
\;|\; 
\etan[T_n] \geq r \; (\forall n)\}, 
\end{equation}
\begin{equation}
a^*_0 \defeq 
\sup\, 
\{\,a\;|\; \zetasupcomp (a) - a \geq r\,\} 
= 
\inf\, 
\{\,a\;|\; \zetasupcomp (a) - a \leq r\,\}. 
\end{equation}
\end{theorem}

\begin{remark}
{\rm 
Note that $\geq$ and $\leq$ in the definition of 
$a^*_0$ can be replaced with $>$ and $<$, 
respectively, because $ \zetasupcomp (a) - a$ is 
a strictly decreasing function. 
}
\end{remark}

\begin{proof}
Suppose that a sequence of tests $\Tvec=\{T_n\}$ 
satisfies $\etainf[\Tvec] \geq r$.  
It then follows from (\ref{eq:eta_and_zeta^c.1}) that 
\begin{align*}
\zetasupcomp[\Tvec] 
&\geq
\sup_a \min \left\{ \zetasupcomp (a), \; \etainf[\Tvec] +a \right\} 
\\
&\geq
\sup_a \min \left\{ \zetasupcomp (a), \; r+a \right\} \\
&\geq
\sup_a\, 
\{\,r+a\;|\; \zetasupcomp (a) \geq r + a\,\} = r+a^*_0, 
\end{align*}
which proves 
\begin{equation}
\label{converse}
B_e^*(r\,|\, \rhovec\,\|\,\sigmavec) 
\geq 
\sup_a \; \min\, \{\zetasupcomp (a), r+a\}
\geq
r + a^*_0 .
\end{equation}

Next, we show that, for arbitrarily given 
$r$, $a$ and $n$, there exists a test $T_n$ 
satisfying 
\begin{equation}
\label{cond_Tn_B*e}
\etan [T_n] \geq r \quad \mbox{and} \quad
\zetancomp [T_n] \leq \max\, \{\zetancomp (a), \, r+a\}. 
\end{equation}
When $\etan (a)\geq r$, 
it is obvious that $T_n\defeq S_n(a)$ 
satisfies this condition.  
When $\etan (a) < r$, let 
\begin{align}
T_n \defeq & I - e^{-n(r-\etan (a))} \; 
S_n^{\,c}(a) \nonumber \\
=& S_n(a) + \{1-e^{-n(r-\etan (a))}\} \; 
S_n^{\,c} (a) .
\label{randomized_test}
\end{align}
In other words, $T_n$ is a randomized test which 
rejects the hypothesis $\rhon$ with probability 
$e^{-n(r-\etan (a))}$ when, and only when, 
the test $S_n(a)$ rejects $\rhon$.  Then we have 
\begin{align*}
\alphan [\Tn] = &
\rhon [e^{-n(r-\etan (a))} S_n^{\,c} (a) ]  \\
= & e^{-n(r-\etan (a))} \alphan (a) = e^{-nr} 
\end{align*}
and 
\begin{align*}
\betancomp [\Tn] 
= & \sigman [e^{-n(r-\etan (a))} S_n^{\,c} (a) ] \\
= &
e^{-n(r-\etan (a))} \betancomp (a) 
\geq e^{-n (r+a)} , 
\end{align*}
where the last inequality follows from 
(\ref{Negative_S^c_n(a).var}).  These are 
rewritten as 
$
\etan[T_n] =  r
$
and 
$
\zetancomp[T_n]
\leq r + a 
$, 
and imply (\ref{cond_Tn_B*e}). 

We have thus shown that for every $r$ and $a$ 
there exists a sequence of tests $\Tvec =\{T_n\}$ 
such that $\etan[T_n] \geq r$ for every $n$ 
and 
\begin{eqnarray*}
\zetasupcomp[\Tvec] &\leq& 
\limsup_{n\rightarrow\infty} 
\; \max \{\zetancomp (a),\, r+a\} \\
&=& \max\, \{\zetasupcomp (a),\, r+a\}, 
\end{eqnarray*}
which leads to
\begin{align}
\nonumber 
B_{e,1}^*(r\,|\, \rhovec\,\|\,\sigmavec) 
&\leq 
\inf_a \; \max\, \{\zetasupcomp (a), \, r+a\} \\
&\leq
\inf_a\, \{r+a\;|\; \zetasupcomp (a) \leq r+a \} 
= r + a^*_0 .
\end{align}
Since $B_{e}^*(r\,|\, \rhovec\,\|\,\sigmavec) 
\leq 
B_{e,1}^*(r\,|\, \rhovec\,\|\,\sigmavec) $ 
is obvious, this together with (\ref{converse}) 
completes the proof. 
\end{proof}

\begin{remark}
{\rm 
The proof of lemma \ref{lemma:eta_and_zeta^c} can be modified, 
using (\ref{formula_limsup}) instead of (\ref{formula_liminf}), 
to yield 
\begin{equation}
\zetainfcomp[\Tvec] 
\geq
\min \,\{ \zetainfcomp (a), \; a+ \etainf[\Tvec] \,\} , 
\end{equation}
whereby we can similarly show that
\begin{align}
\hat{B}_e^*(r\,|\, \rhovec\,\|\,\sigmavec)
&=  \hat{B}_{e,1}^*(r\,|\, \rhovec\,\|\,\sigmavec) \\
&= \sup_a \; \min\,\{\zetainfcomp (a), \, r+a\} \\
&= \inf_a \; \max\, \{\zetainfcomp (a), \, r+a\} \\
&= r + \hat{a}^*_0, 
\end{align}
where
\begin{align}
\hat{B}_e^*(r\,|\, \rhovec\,\|\,\sigmavec) 
&\defeq 
\inf_{\Tvec}\, \{\,\zetainfcomp [\Tvec] 
\;|\; \etainf[\Tvec] \geq r\} , \\
\hat{B}_{e,1}^*(r\,|\, \rhovec\,\|\,\sigmavec) 
&\defeq 
\inf_{\Tvec}\, \{\,\zetainfcomp [\Tvec] 
\;|\; 
\etan[T_n] \geq r \; (\forall n)\}, 
\end{align}
and
\begin{equation}
\hat{a}^*_0 \defeq 
\sup\, 
\{\,a\;|\; \zetainfcomp (a) - a \geq r\,\} 
= 
\inf\, 
\{\,a\;|\; \zetainfcomp (a) - a \leq r\,\}. 
\end{equation}
}
\end{remark}

\medskip

Now let us demonstrate how Han's formula 
(\ref{Han_B^*}) is derived from Theorem~\ref{thm:Be*}. 

\begin{lemma}
It always holds that 
\begin{equation}
\label{zepsupcomp&etasup}
\zetasupcomp (a) \leq a + \etasup (a) 
\end{equation}
for any $a$, and  
if $a$ is a decreasing point of 
$\zetasupcomp$ in the sense that 
$\zetasupcomp (a-\varepsilon) > 
\zetasupcomp (a+\varepsilon)$ 
for any $\varepsilon >0$, then we also have 
\begin{equation}
\label{zepsupcomp&etainf}
\zetasupcomp(a+0) \geq a + \etainf (a+0).
\end{equation}
\end{lemma}

\begin{proof}
The inequalities are immediate from 
(\ref{Negative_S^c_n(a).var}) and 
(\ref{eq:eta_and_zeta^c.2}), respectively. 
\end{proof}

\begin{corollary}
\label{cor:Han_B^*}
It always holds that 
\begin{eqnarray}
\nonumber
B_{e}^*(r\,|\, \rhovec\,\|\,\sigmavec) 
&\leq& 
\inf_a \, \Bigl(
a+ \max\, \{\etasup (a) ,\, r\} \Bigr) \\
&= &
\inf_a\, \Bigl\{ a+ \etasup (a) + 
[r-\etasup (a)]_+\Bigr\}, 
\label{ineq1:Han_B^*}
\end{eqnarray}
while we have
\begin{eqnarray}
\nonumber
B_{e}^*(r\,|\, \rhovec\,\|\,\sigmavec) 
&\geq &
\inf_a \, \Bigl(
a+ \max\, \{\etainf (a) ,\, r\} \Bigr) \\
&= &
\inf_a\, \Bigl\{ a+ \etainf (a) + 
[r-\etainf (a)]_+\Bigr\} 
\label{ineq2:Han_B^*}
\end{eqnarray}
if 
\begin{equation}
\label{cond1:Han_B^*}
r 
\leq \zetasupcomp (-\infty) 
-
\sup\, \{ a\,|\, \zetasupcomp (-\infty) = 
\zetasupcomp (a)\,\} . 
\end{equation}
In particular, if 
the limit 
$\eta (a)  \defeq 
\lim_{n\rightarrow\infty} 
\etan (a) $ exists for all $a$ 
and if 
\begin{equation}
\label{cond2:Han_B^*}
\zetasupcomp (-\infty) = \infty 
\quad \mbox{or} \quad 
\{ a\,|\, \zetasupcomp (-\infty) = 
\zetasupcomp (a)\,\}
=\phi, 
\end{equation}
then Han's formula (\ref{Han_B^*}) is valid 
for all $r$. 
\end{corollary}

\begin{proof}
Since the first inequality is immediate from 
Theorem~\ref{thm:Be*} and (\ref{zepsupcomp&etasup}), 
we only prove the second one.  Invoking Theorem~\ref{thm:Be*} 
again, it suffices to show 
that, for any $r$ satisfying (\ref{cond1:Han_B^*}),  
\begin{equation}
\label{ineq:proof_Han_B^*}
\inf_a 
\max\, \{\zetasupcomp (a), \, r+a\} 
\geq 
\inf_{a} \, \Bigl(
a+ \max\, \{\etainf (a) ,\, r\} \Bigr) .
\end{equation}
Define  
\[ 
b_0 \defeq \sup\, 
\{ a\,|\, \zetasupcomp (-\infty) = 
\zetasupcomp (a)\,\} , 
\]
including the case when $\{ a\,|\, \zetasupcomp (-\infty) = 
\zetasupcomp (a)\,\} =\phi$ and $b_0=-\infty$, 
and let $a$ be an arbitrary number satisfying 
$a > b_0$ .  Then we have 
$\zetasupcomp (-\infty) > \zetasupcomp (a)$, 
which means that there exists a number $b$ such that 
$b < a$ and $\zetasupcomp (b) > \zetasupcomp (a) $.  
The supremum of such numbers $b$, denoted by $\bar{b}=\bar{b}(a)$, 
satisfies $\bar{b}\leq a$ and 
$
\zetasupcomp (\bar{b} -\varepsilon) > \zetasupcomp (a) 
\geq \zetasupcomp (\bar{b} +\varepsilon)
$
for every $\varepsilon >0$, which implies that 
$\bar{b}$ is a 
decreasing point of $\zetainfcomp$ 
and $\zetasupcomp (a) \geq \zetasupcomp (\bar{b} +0)$. 
Hence we have  
\begin{align*}
\max\, \{\zetasupcomp (a), \, r+a\} 
&\geq 
\max\, \{\zetasupcomp (\bar{b} +0), \, r+\bar{b}\} \\
&\geq 
\max\, \{\etainf (\bar{b} +0) + \bar{b}, \, r+\bar{b}\} \\
&= 
\lim_{\varepsilon\downarrow 0} 
\Bigl( \bar{b}+\varepsilon + \max\, \{\etainf (\bar{b} +\varepsilon), r\}
\Bigr) \\
&\geq 
\inf_{b} \, \Bigl(
b+ \max\, \{\etainf (b) ,\, r\} \Bigr) , 
\end{align*}
where the second inequality follows from (\ref{zepsupcomp&etainf}), 
and therefore
\begin{equation}
\label{ineq1:proof_Han_B^*}
\inf_{a\,:\,a>b_0} 
\max\, \{\zetasupcomp (a), \, r+a\} 
\geq 
\inf_{a} \, \Bigl(
a+ \max\, \{\etainf (a) ,\, r\} \Bigr) .
\end{equation}
This proves (\ref{ineq:proof_Han_B^*}) when 
$b_0 =-\infty$.  In the case when $b_0 > -\infty$, we have
\begin{align}
\nonumber 
\inf_{a\,:\,a<b_0} 
\max\, \{\zetasupcomp (a), \, r+a\} 
&\geq \inf_{a\,:\,a<b_0} \zetasupcomp (a) \\
\nonumber
&= \zetasupcomp (-\infty) \\
\label{ineq2:proof_Han_B^*}
&\geq \max\,\{ \zetasupcomp (b_0), \, r+b_0 \},
\end{align}
where the last inequality follows from 
the nonincreasing property of $\zetasupcomp$ and 
(\ref{cond1:Han_B^*}) , and in addition we have
\begin{align}
\nonumber
\max\,\{ \zetasupcomp (b_0), \, r+b_0 \} 
&\geq \max\,\{ \zetasupcomp (b_0+0), \, r+b_0 \} \\
\nonumber
&= \lim_{\varepsilon\downarrow 0} 
\max\,\{ \zetasupcomp (b_0+\varepsilon), \, r+b_0+\varepsilon \} \\
\label{ineq3:proof_Han_B^*}
&\geq 
\inf_{a\,:\,a>b_0}\, \max\,\{ \zetasupcomp (a), \, r+a \}  .
\end{align}
Now the desired inequality (\ref{ineq:proof_Han_B^*}) 
follows from (\ref{ineq1:proof_Han_B^*}), (\ref{ineq2:proof_Han_B^*}) 
and (\ref{ineq3:proof_Han_B^*}). 
\end{proof}

\begin{remark}
\label{remark:Han_B^*}
{\rm 
The first condition in (\ref{cond2:Han_B^*}) 
in terms of $\zetasupcomp$ 
is weaker than the original condition 
(\ref{unbounded_zetainfcomp}) in terms of $\zetainfcomp$. 
 The second condition  in (\ref{cond2:Han_B^*}) 
means that for any number $a$ there always exists 
a number $b<a$ such that $\zetasupcomp (b) > 
\zetasupcomp (a)$.  The second is not implied by the first, 
since there may be a number $a$ such that 
$\zetasupcomp (b) = \infty$ for all $b >a$. 
}
\end{remark}

\begin{remark}
{\rm 
It is easy to see that 
if the condition  (\ref{cond1:Han_B^*}) 
is not satisfied then 
\[
B_{e}^*(r\,|\, \rhovec\,\|\,\sigmavec) 
= \zetainfcomp (-\infty) .
\] 
}
\end{remark}

\begin{example}
{\rm 
Consider the following classical hypothesis testing problem: 
$\cXn = \{x_0, x_1\}$ for all $n$ on which probability 
distributions $\rhon$ and $\sigman$ are defined by 
\begin{gather}
 \rhon (x_0) = e^{-n b_n}, \quad \rhon (x_1) = 1-e^{-n b_n}, \\
\sigman (x_0) = e^{-n c}, \quad \sigman (x_1) = 1-e^{-n c}, 
\end{gather}
where $b_n$ is a positive sequence obeying 
$\lim_{n\rightarrow\infty} b_n = \infty$ and 
$c$ is a positive constant. Then 
the limits $\eta (a) = \lim_{n\rightarrow\infty}\etan (a)$ and 
$\zeta^{\, c}(a) = \lim_{n\rightarrow\infty} \zetan^{\, c} (a)$ 
exist for all $a$ and satisfy
\[
\eta (a) = \left\{
\begin{array}{ccc}
0 &\mbox{if}& a>0 \\
\infty &\mbox{if}& a\leq 0
\end{array}
\right. 
\quad \mbox{and}\quad 
\zeta^{\, c} (a) = \left\{
\begin{array}{ccc}
0 &\mbox{if}& a>0 \\
c &\mbox{if}& a\leq 0 ,
\end{array}
\right.
\]
where we have chosen $S_n (a) = 
\{ \rhon - e^{na}\sigman > 0\}$ in 
(\ref{Def_S_n(a).general}).  Note also that 
$\infD (\sigmavec\,\|\,\rhovec) = 
\supD (\sigmavec\,\|\,\rhovec) = 0$. 
It is then immediate from Theorem~\ref{thm:Be*} 
that
\[
B_e^*(r\,|\, \rhovec\,\|\,\sigmavec) 
= 
\left\{
\begin{array}{ccc}
c &\mbox{if} &r\geq c \\
r &\mbox{if} & 0\leq r\leq c \\
0  &\mbox{if} & r\leq 0 ,
\end{array}
\right.
\]
while we have
\[
\inf_a\, \Bigl\{ a+ \eta (a) + 
[r-\eta (a)]_+\Bigr\} 
= 
\left\{
\begin{array}{ccc}
r &\mbox{if} & r\geq 0 \\
0 &\mbox{if} & r\leq 0 .
 \end{array}
\right.
\]
Since 
$\zeta^{\, c} (-\infty) = c$ and 
$\sup\, \{ a\,|\, \zeta^{\, c} (-\infty) = 
\zeta^{\, c} (a)\,\}
= 0$, the condition (\ref{cond1:Han_B^*}) 
for validity of Han's formula turns out 
to be $r\leq c$,  which just explains the above 
situation.  
}
\end{example}

\begin{remark}
{\rm 
In the classical i.i.d.\ case,  Han and Kobayashi \cite{Han-Kobayashi} 
(see also \cite{Nakagawa-Kanaya}) 
obtained a compact expression for
$B_e^*(r\,|\, \rhovec\,\|\,\sigmavec)$ in the form
\begin{equation}
B_e^*(r\,|\, \rhovec\,\|\,\sigmavec) 
= \min_{\tau\,:\, D(\tau\,\|\,\rho) \leq r} 
\left\{
D(\tau\,\|\,\sigma)+r- D(\tau\,\|\,\rho) 
\right\} 
\label{eq:han_kob}
\end{equation}
with noting that the RHS can be represented as 
$\displaystyle \min_{\tau\,:\,D(\tau\,\|\,\rho) \geq r} 
D(\tau\,\|\,\sigma)$ 
when $r$ is sufficiently near $D(\sigma \,\|\,\rho)$. 
We also have an expression in the form\footnote{%
To the authors' knowledge, this type of expression 
for $B_e^*$ 
first appeared in \cite{Ogawa-Nagaoka} even for the classical case. 
}
\begin{equation}
B_e^*(r\,|\, \rhovec\,\|\,\sigmavec) 
= \max_{\theta \leq -1}\frac{(1+\theta)r +\psi(\theta)}{\theta}, 
\label{eq:han_kob_var}
\end{equation}
where $\psi(\theta) = \log \sum_x \rho(x)^{1+\theta}\sigma(x)^{-\theta}$ 
is the same function as defined in Remark~\ref{remark:Be_quant}. 
These expressions can be derived by applying large deviation theorems to 
 Theorem~\ref{thm:Be*} (or to (\ref{Han_B^*}) as in \cite{Han_book, 
Han_test}) 
(cf.\ Remark~\ref{remark:Be_quant}).  
For the quantum i.i.d.\ case, it was shown in \cite{Ogawa-Nagaoka} 
that inequality (\ref{eq:OgaNag}), with $\rho$ and $\sigma$ exchanged, 
yields a lower bound on $B_e^*(r\,|\, \rhovec\,\|\,\sigmavec)$ 
in the same form as the RHS of (\ref{eq:han_kob_var}) 
except that the range of $\max$ is restricted to 
$-2 \leq \theta \leq -1$ (; see \cite{Nagaoka:strong} for a simple 
derivation). 
 This restriction has been relaxed to $\theta \leq -1$ 
 just as (\ref{eq:han_kob_var}) by \cite{Hayashi:hypo_quant} 
 and \cite{Hayashi:book} (section~3.4)\footnote{%
 Note that $B^*(r\,|\rho\,\|\,\sigma)$ in \cite{Hayashi:book} 
corresponds to our $B_e^* (r\,|\,\sigmavec\,\|\,\rhovec )$. 
 }.
 Some further results on $B_e^*(r\,|\, \rhovec\,\|\,\sigmavec)$ 
 are also found in \cite{Hayashi:book} 
 including a quantum extension of (\ref{eq:han_kob_var}) 
 (not a bound but an identity) 
 in terms of a variant of $\psi(\theta)$ which is 
 defined in a limiting form\footnote{%
 This result has not appeared in the original Japanese edition of 
\cite{Hayashi:book}.}
 }.
\end{remark}

Before concluding this section, we introduce the dual of 
$B_e^{*}(r\,|\, \rhovec\,\|\,\sigmavec) $
by
\begin{align}
\nonumber
B_e^{**}(r\,|\, \rhovec\,\|\,\sigmavec) 
&\defeq 
\sup_{\Tvec}\, \{\, 
\etainf [\Tvec]  \; |\;
\zetasupcomp[\Tvec] \leq r\} \\
&= 
\sup\, \{\,r'\,|\, B_e^* (r'\,|\, \rhovec\,\|\,\sigmavec) \leq r
\} 
\label{Def_Be**}
\end{align}
and provide this with a general characterization, 
which will be applied to the source coding 
problem in the next section.

\begin{theorem}
\label{thm:B**e}
We have
\begin{equation}
B_e^{**}(r\,|\, \rhovec\,\|\,\sigmavec) 
=
r - a^{**}_0, 
\end{equation}
where
\begin{equation}
a^{**}_0 \defeq \inf\, \{\,a\,|\,\zetasupcomp (a) \leq r \} 
= \sup\, \{\,a\,|\,\zetasupcomp (a) > r\}. 
\end{equation}
\end{theorem}

\begin{proof}
Obvious from the following equivalence:
\begin{align*}
B_e^* (r' \,|\, \rhovec\,\|\,\sigmavec) \leq r 
&\Leftrightarrow \; 
\sup_a \, \min\,\{\zetasupcomp (a), \, r' +a\} \leq r \\
&\Leftrightarrow \; 
\forall a \; 
\left(
a> r-r' \;\; \Rightarrow \zetasupcomp (a) \leq r \right) \\
&\Leftrightarrow \; 
r'\leq r- a^{**}_0.
\end{align*}
\end{proof}

\section{Application to classical fixed-length source coding}
\label{sec:source}

Let $\rhovec=\{\rhon\}_{n=1}^\infty$ be a classical general 
source; i.e., let $\rhon$ be a probability distribution 
$P_{\Xn}$ on a finite or countably infinite set $\cXn$ for 
each $n$.  A (possibly stochastic) fixed-length coding system for this 
source 
is generally represented by a sequence $\Phivec = \{\Phin\}_{n=1}^\infty$ 
of $\Phin = (\cYn, \Fn, \Gn)$, where $\cYn$ is a finite set, $\Fn = 
\{\Fn (y\,|\,x)\}$ is 
a channel from $\cXn$ to $\cYn$ representing an encoder, 
and $\Gn=\{\Gn(x\,|\,y)\}$ is a channel from 
$\cYn$ to $\cXn$ representing a decoder. The size and the 
error probability of $\Phin$ are respectively defined 
by $|\Phin|\defeq |\cYn|$ and 
\[\gamman[\Phin]\defeq 
1- \sum_{x\in\cXn}\sum_{y\in\cYn} \rhon(x) \Fn (y\,|\,x) \Gn(x\,|\,y).
\]

Now, let $\sigman$ be the counting measure on $\cXn$;  
i.e., $\sigman [\Tn] = \sum_{x\in\cXn} \Tn (x)$.  Then 
the source coding problem for $\rhon$ can be reduced to 
the generalized hypothesis 
testing problem for $\{\rhon, \sigman\}$ as follows. 
For an arbitrary coding system $\Phin$, 
a test $\Tn$ is defined by 
\begin{equation}
\Tn (x) = \sum_{y\in\cYn} \Fn (y\,|\,x) \Gn(x\,|\,y) ,
\end{equation}
which satisfies 
\begin{equation}
\label{gamma=alpha} 
\gamman[\Phin] = 1-\rhon[\Tn] = \alphan[\Tn] 
\end{equation}
and
\begin{align}
|\Phin | 
= &
|\cYn| = \sum_{y\in\cYn} 1
\geq \sum_{y\in\cYn} \sum_{x\in\cXn} \Fn(y\,|\,x) \Gn(x\,|\,y) \nonumber \\
= &
\sigman[\Tn]  
= \betan[\Tn].
\label{Phi_geq_beta}
\end{align}
Conversely, for an arbitrary deterministic (i.e., $\{0,1\}$-valued) test 
$\Tn$ 
we can construct 
a coding system $\Phin=(\cYn, \Fn, \Gn)$ satisfying 
$\gamman[\Phin] = \alphan[\Tn]$ and 
$|\Phin | =  \betan[\Tn]$ 
by setting $\cYn= \{x\in\cXn\,|\, \Tn(x)=1\}$ 
and 
$\Fn(y\,|\,x) = \Gn(x\,|\,y)= 1$ if  
$y=x\in\cYn$. 
Noting that the direct (achievability) parts of 
Theorems~\ref{Th:SteinSpec} and \ref{thm:Be} 
have been shown by using only deterministic tests 
(by setting $S_n(a)$ to be $\{\rhon -e^{na}\sigman >0\}$ 
or $\{\rhon -e^{na}\sigman \geq 0\}$)
and that 
$\limsup_{n\rightarrow\infty}
\frac{1}{n}\log |\Phin | = 
- \zetainf[\Tvec]$ when $|\Phin | = \betan[\Tn]$, 
we immediately obtain the following identities:

\begin{theorem}
\label{thm:R}
We have
\begin{align}
\nonumber 
R(\varepsilon\, |\,\rhovec) 
&\defeq 
\inf_{\Phivec}\, \{\,
\limsup_{n\rightarrow\infty} \frac{1}{n} \log |\Phin|
\;|\; 
\limsup_{n\rightarrow\infty} \gamman[\Phin] \leq\varepsilon \} \\
\label{R=-B:epsilon.1}
&=
- B (\varepsilon\,|\,\rhovec\,\|\,\sigmavec) 
=
- \infD (\varepsilon\,|\,\rhovec\,\|\,\sigmavec) \\
\label{R=-B:epsilon.2}
&= \inf\, \big\{a\,\big|\, \limsup_{n\rightarrow\infty} 
P_{\Xn}\bigl\{-\frac{1}{n}\log P_{\Xn}(\Xn) \ggeq 
a\bigr\}\leq\varepsilon\big\}, 
\end{align}
\begin{align}
\nonumber 
R^\dagger(\varepsilon\, |\,\rhovec) 
&\defeq 
\inf_{\Phivec}\, \{\,
\limsup_{n\rightarrow\infty} \frac{1}{n}  \log |\Phin |
\;|\; 
\liminf_{n\rightarrow\infty} \gamman [\Phin] < \varepsilon \} \\
\label{R+=-B+:epsilon.1}
&= 
- B^\dagger (\varepsilon\,|\,\rhovec\,\|\,\sigmavec)
= 
- \supD (\varepsilon\,|\,\rhovec\,\|\,\sigmavec) \\
\label{R+=-B+:epsilon.2}
&= \inf\, \big\{a\,\big|\, \liminf_{n\rightarrow\infty} 
P_{\Xn}\bigl\{-\frac{1}{n}\log P_{\Xn}(\Xn) \ggeq a\bigr\} < 
\varepsilon\big\}, 
\end{align}
\begin{align}
R(\rhovec) 
\nonumber 
&\defeq R(0\, |\,\rhovec) \\
\nonumber 
& = 
\inf\,\{R\,|\, 
\exists \Phivec, \; 
\limsup_{n\rightarrow\infty} \frac{1}{n} \log |\Phin | \leq R 
\;\;\mbox{and}\;
\lim_{n\rightarrow\infty} \gamman [\Phin] =0 \} \\
\label{R=-B:0.1}
&= 
- B (\rhovec\,\|\,\sigmavec)
= 
- \infD (\rhovec\,\|\,\sigmavec) \\
\label{R=-B:0.2}
&= \supH(\rhovec)
\defeq 
\mbox{p-}\limsup_{n\rightarrow\infty}\, 
\bigl\{-\frac{1}{n}\log P_{\Xn} (\Xn)\bigr\}, 
\end{align}
\begin{align}
\nonumber 
R^\dagger (\rhovec) 
&\defeq
R^\dagger (1\, |\,\rhovec)  \\
\nonumber 
&=
\sup\,\{R\,|\, 
\forall \Phivec, \; 
\mbox{if}\;\;
\limsup_{n\rightarrow\infty} \frac{1}{n} \log |\Phin | \leq R 
\;\;\mbox{then}\;
\lim_{n\rightarrow\infty} \gamman [\Phin] =1 \} \\
\label{R+=-B+:1.1}
&= 
- B^\dagger (\rhovec\,\|\,\sigmavec)
= 
- \supD (\rhovec\,\|\,\sigmavec) \\
\label{R+=-B+:1.2}
&= \infH(\rhovec)
\defeq 
\mbox{p-}\liminf_{n\rightarrow\infty}\, 
\bigl\{-\frac{1}{n}\log P_{\Xn} (\Xn)\bigr\} ,
\end{align}
and 
\begin{align}
\nonumber 
R_e (r\,|\,\rhovec)
&\defeq
\inf_{\Phivec}\, \big\{\, 
\limsup_{n\rightarrow\infty} \frac{1}{n} \log |\Phin |
\;\big|\; 
\liminf_{n\rightarrow\infty} 
\bigl\{- \frac{1}{n} \log\gamman[\Phin] \bigr\} \geq r \big\} \\
\label{Re=-Be.1}
&= 
- B_e (r\,|\, \rhovec\,\|\,\sigmavec) \\
\label{Re=-Be.2}
&= 
\sup_{a}\,\{a-\sigmainf(a) \,|\, \sigmainf(a) < r\}, 
\end{align}
where 
\begin{align}
& \sigmainf(a) \defeq \etainf(-a)  \nonumber\\
=& 
\liminf_{n\rightarrow\infty} 
\,  -\frac{1}{n} \log P_{\Xn} 
\bigl\{-\frac{1}{n}\log P_{\Xn}(\Xn) \ggeq a \bigr\} .
\label{def:sigmainf}
\end{align}
\end{theorem}

\begin{remark}
{\rm 
$R(\rhovec)$ is the optimal compression rate 
with asymptotically vanishing error probability 
and Equation (\ref{R=-B:0.2}), which was originally 
shown in \cite{HanVerdu_output}, means that it always 
equals the spectral sup-entropy rate $\supH(\rhovec)$. 
The source $\rhovec$ is said to have the strong converse 
property when 
$ \limsup_{n\rightarrow\infty} \frac{1}{n} \log |\Phin | 
< R(\rhovec) 
$
implies 
$
\lim_{n\rightarrow\infty} \gamman [\Phin] =1
$,
or equivalently when $R(\rhovec) = R^\dagger (\rhovec)$.  
As was pointed out in \cite{HanVerdu_output}, 
this property is equivalent to 
$\supH(\rhovec) = \infH (\rhovec)$, 
which is now obvious from (\ref{R=-B:0.2}) and (\ref{R+=-B+:1.2}). 
Equation (\ref{R=-B:epsilon.2}) is found in \cite{SteinbergVerdu}, and 
(\ref{Re=-Be.2}) in
\cite{Han_source}.  
Although the use of the symbol $\sigma$ for different notions, 
for the counting measures and for the function (\ref{def:sigmainf}), 
 may be a little confusing, it will be helpful for comparing our results 
with 
those of \cite{Han_source}.
}
\end{remark}

Next, let us turn to the following quantity:
\begin{align}
&R^*_e (r\,|\,\rhovec)\nonumber \\
\defeq &
\inf_{\Phivec} \big\{ 
\limsup_{n\rightarrow\infty} \frac{1}{n} \log | \Phin |
\big| 
\limsup_{n\rightarrow\infty} 
\bigl\{- \frac{1}{n} \log (1- \gamman[\Phin] ) \bigr\} \leq r \big\} .
\end{align}
Han \cite{Han_book, Han_source} proved that for any $r\geq 0$ 
\begin{equation}
\label{Han_R^*}
R^*_e (r\,|\,\rhovec)
=
\inf\,\bigl\{h\geq 0\,\big|\,
\inf_{a} \, \{\sigma^* (a) 
+ [a-\sigma^*(a) -h]^+\}\leq r
\bigr\} 
\end{equation}
under the assumption that the 
following limit exists for all $a$:
\[
\sigma^*(a) 
\defeq 
\lim_{n\rightarrow\infty} \,  -\frac{1}{n} \log P_{\Xn} 
\bigl\{-\frac{1}{n}\log P_{\Xn}(\Xn) \leq a \bigr\} . 
\] 
A general formula for $R^*_e (r\,|\,\rhovec)$ which needs no 
additional assumption is given below. 

\begin{theorem} 
\label{thm:R*e} 
For any $r\geq 0$ we have 
\begin{equation}
\label{formula:R*e}
R^*_e (r\,|\,\rhovec) = 
\max \, \{ b_0 - r, \, 0 \}, 
\end{equation}
where 
\[
b_0 \defeq 
\sup \, \{a\,|\, \sigmasupast (a) > r\} = 
\inf \, \{a\,|\, \sigmasupast (a) \leq r\} 
\]
and
\begin{align*}
\sigmasupast (a) 
&\defeq 
\limsup_{n\rightarrow\infty} \,  -\frac{1}{n} \log P_{\Xn} 
\bigl\{-\frac{1}{n}\log P_{\Xn}(\Xn) \lleq a \bigr\} . 
\end{align*}
\end{theorem}

\begin{proof}
Since 
\[ 
\sigmasupast (a) 
= 
\limsup_{n\rightarrow\infty} \,  -\frac{1}{n} \log \rhon 
\bigl[ \{\rhon - e^{-na}\sigman \geq 0 \} \bigr]  
= \etasupcomp (-a) , 
\]
it follows from Theorem~\ref{thm:B**e}, with $\rhovec$ and $\sigmavec$ 
exchanged, that 
\[
B^{**}_e (r\,|\,\sigmavec\,\|\,\rhovec ) 
=  r-b_0 . 
\]
Hence (\ref{formula:R*e}) is equivalent to
\begin{equation} 
R^*_e (r\,|\,\rhovec) = \max\,\{- B^{**}_e (r\,|\,\sigmavec\,\|\,\rhovec ) 
, \, 0 \} .
\end{equation} 
Here it is easy to see LHS $\geq$ RHS 
from (\ref{gamma=alpha}), (\ref{Phi_geq_beta}) and 
\begin{align*}
&- B^{**}_e (r\,|\,\sigmavec\,\|\,\rhovec ) 
=
- \sup_{\Tvec}\, \{\, 
\zetainf [\Tvec]  \; |\;
\etasupcomp[\Tvec] \leq r\} \\
& = \inf_{\Tvec} \, \{\limsup_{n\rightarrow\infty} 
\frac{1}{n} \log \betan [\Tn] \, |\, 
\limsup_{n\rightarrow\infty} 
-\frac{1}{n} \log (1-\alphan [\Tn] ) \leq r \}. 
\end{align*}
Let us show the converse inequality LHS $\leq$ RHS. 
Since 
\[ 
- B^{**}_e (r\,|\,\sigmavec\,\|\,\rhovec ) 
=
\inf\,\{ r'\, |\, B^*_e (-r'\,|\,\sigmavec\,\|\,\rhovec )\leq r\}, 
\]
it is sufficient to show that 
\[ R^*_e (r\,|\,\rhovec) \leq r'
\quad\mbox{if}\quad 
B^*_e  (-r' \,|\,\sigmavec\,\|\,\rhovec )\leq r
\;\;\mbox{and}\;\; r' >0.
\] 
This is equivalent to the 
proposition that for any 
$r'>0$ there exists a sequence of codes 
$\Phivec$ such that 
\begin{gather*}
\limsup_{n\rightarrow\infty} \frac{1}{n} \log |\Phin| \leq r', 
\quad\mbox{and} \\
\limsup_{n\rightarrow\infty} -\frac{1}{n} \log (1-\gamma_n [\Tn]) \leq 
B^*_e(-r' \,|\,\sigmavec\,\|\,\rhovec ) .
\end{gather*}
This proposition follows if 
the infimum of
\[
B^*(-r' \,|\,\sigmavec\,\|\,\rhovec ) 
=
\inf_{\Tvec} \, 
\{\etasupcomp [\Tvec]\,|\, 
\zetainf [\Tvec] \geq -r'\}
\]
can be attained by a sequence of 
{\rm deterministic} tests $\Tvec$ when $r'>0$. 
Recalling the proof of Theorem~\ref{thm:Be*} 
and applying it to the present situation, 
it suffices to show that for any $a\in\bR$, $r'>0$, $\delta>0$ and 
for any sufficiently large $n$ there 
exists  a deterministic test satisfying 
\begin{equation}
\label{cond_Tn_B*e_var}
\zetan [T_n] \geq -r'-\delta \quad \mbox{and} \quad
\etancomp [T_n] \leq \max\, \{\etancomp (a), \, -r'-a\}, 
\end{equation}
which corresponds to (\ref{cond_Tn_B*e}) with 
a slight modification.  
Let $S_n(a)$ be chosen to be deterministic in 
(\ref{Def_S_n(a).general}) and identify it with 
its acceptance region (e.g.,   
$S_n(a) = \{x\in\cXn \,|\, \rhon(x) - e^{na} \sigman (x) 
>0\}$). 
It is then obvious that $T_n\defeq S_n(a)$ 
satisfies (\ref{cond_Tn_B*e_var}) if $\zetan (a)\geq -r'$. 
Suppose $\zetan (a) < -r'$, which means $\sigman [S_n(a)] 
= |S_n(a)| > e^{nr'}$. 
Then there exists a subset $\Tn\subset S_n (a)$, 
which is regarded as a deterministic test, satisfying 
\[ \sigma [\Tn ] = |\Tn | = \lceil e^{nr'}\rceil 
\quad\mbox{and}\quad
\rhon[\Tn ] \geq \frac{ \lceil e^{nr'}\rceil}{ |S_n (a) |} \; 
\rhon [S_n (a)] . 
\]
Using (\ref{Positive__S_n(a)}) we have $\rhon[\Tn ]\geq 
e^{n(a+r')}$, and $\Tn$ satisfies (\ref{cond_Tn_B*e_var}). 
\end{proof}

Now we can see that Han's formula (\ref{Han_R^*}) 
is very near to the true general formula.  Actually, 
if $\sigma^*(a)$ is simply replaced with $\sigmasupast (a)$, 
it becomes equivalent to (\ref{formula:R*e}) as follows.  
Noting that $\sigmasupast (a)$ is monotonically nonincreasing, we have 
\begin{align*}
& 
\inf_{a} \, \{\sigmasupast (a) 
+ [a-\sigmasupast (a) -h]^+\}\leq r \\
\Leftrightarrow \quad & 
\inf_{a} \, \max\,\{ 
\sigmasupast (a) , 
a- h \}\leq r \\
\Leftrightarrow \quad & 
\forall\delta >0, \; \exists a,  \; 
\sigmasupast (a) \leq r+\delta 
\;\;\mbox{and}\;\;  
a-h\leq r+\delta \\
\Leftrightarrow \quad & 
h \geq \; \sup_{\delta >0} \; \inf_{a}\; \{ a - (r + \delta) \,|\, 
\sigmasupast (a) \leq r+\delta \,\} \\
\Leftrightarrow \quad & 
h \geq \; \sup_{\delta >0} \; \sup_{a}\; \{ a - (r + \delta) \,|\, 
\sigmasupast (a) > r+\delta \,\} \\
\Leftrightarrow \quad & 
h \geq \; \sup_{a} \; \{ a - r \,|\, 
\exists\delta >0, \; 
\sigmasupast (a) > r+\delta \,\} \\
\Leftrightarrow \quad & 
h \geq \; \sup_{a} \; \{ a - r \,|\, 
\sigmasupast (a) > r\,\}  = b_0 - r, 
\end{align*}
and therefore
\begin{align*}
&\inf\,\bigl\{h\geq 0\,\big|\,
\inf_{a} \, \{\sigma^* (a) 
+ [a-\sigmasupast(a) -h]^+\}\leq r
\bigr\} \\
=& 
\max \, \{ b_0 - r, \, 0 \}.
\end{align*}

\begin{remark} 
{\rm 
Iriyama (and Ihara) \cite{Iriyama1, Iriyama2} obtained 
other forms of general formulas 
for 
$R_e (r\,|\,\rhovec)$ and $R^*_e (r\,|\,\rhovec)$ 
from 
a different point of view. 
}
\end{remark}

\begin{remark}
{\rm 
Although we have treated 
only classical source coding here, 
extension to some quantum settings is actually possible. 
Of the two major coding schemes proposed  
for the quantum pure sate source coding, 
namely visible coding and blind coding, 
the former is less restrictive and hence needs 
in general 
a more careful or stronger argument than the latter when 
showing the converse part of a 
theorem concerning a limit on all possible 
codes.  The situation is reversed 
when showing the direct (achievability) part.  
It is easy to see that 
the direct parts of Theorems~\ref{thm:R} and~\ref{thm:R*e}
are straightforwardly extended to visible coding,
and that the direct part only of the arguments concerning 
$R(\vec{\rho})$ and $R_e(r|\vec{\rho})$ in Theorems~\ref{thm:R}
is applicable to blind coding,
while it is not clear 
whether other bounds in Theorems~\ref{thm:R} and~\ref{thm:R*e} 
are achievable for blind coding.
On the other hand, it has been shown in \cite{Hayashi:quant_source} 
that the inequality 
\[ 
\gamman[\Phin] + e^{na} |\Phin | \geq 
\rhon [\{\rhon - e^{na}\sigman \lleq 0\}], 
\]
which follows from (\ref{Optimal_S_n(a).var}), 
(\ref{gamma=alpha}) and (\ref{Phi_geq_beta}), 
can be extended to visible coding just in the 
same form. Since the converse parts of our theorems 
are direct consequences of this inequality, they are 
extended to visible coding, and hence to 
blind coding as well. We thus have the same formula 
as Theorem~\ref{thm:R} for both visible 
and blind coding, and Theorem~\ref{thm:R*e} for 
visible coding. 
See \cite{Hayashi:quant_source} for details. 
Hayashi \cite{Hayashi:entanglement}
showed that these values 
$R^\dagger(\epsilon|\vec{\rho})$, $R^\dagger(\vec{\rho})$, 
$R_e(r|\vec{\rho})$
have other operational meaning.
He also treated these values when the quantum information source is 
given by the thermal state of Hamiltonian with interaction.
That is, using this discussion, we can treat the bounds
$R^\dagger(\epsilon|\vec{\rho})$, $R^\dagger(\vec{\rho})$, 
$R_e(r|\vec{\rho})$ in this case.
}
\end{remark}
\begin{remark}\rm
Recently, Hayashi\cite{New} clarified the relation 
between $R(\epsilon|\vec{\rho})$ and $R^{\dagger}(\epsilon|\vec{\rho})$
from a wider view point.
\end{remark}

\section{Concluding remarks}
\label{sec:conclusion}

We have demonstrated that the 
information-spectrum analysis made by Han for the classical 
hypothesis testing for simple hypotheses, 
together with the fixed-length source coding, 
can be naturally extended to a unifying framework 
including both the classical and quantum generalized 
hypothesis testing. 
The generality of theorems and the 
simplicity of proofs have been thoroughly pursued 
and have yielded some improvements of the original classical results. 

The significance of our results for 
quantum information theory is not so clear at present, 
since our knowledge of 
the asymptotic behavior of the quantum information 
spectrum $\rhon[\{\rhon - e^{na}\sigman > 0\}] 
= \Tr (\rhon \{\rhon - e^{na}\sigman > 0\})$ is 
insufficient even for the i.i.d.\ case 
$\rhon =\rho^{\otimes n}, \sigman =\sigma^{\otimes n}$. 
Therefore we cannot obtain compact and computable representations 
of information-spectrum quantities.  Nevertheless, the fact that 
the asymptotic characteristics of quantum hypothesis 
testing are represented in terms of the information spectrum 
seems to suggest the importance of studying 
quantum information theory from the information-spectrum viewpoint.  
An attempt in this direction is found in 
\cite{Hayashi-Nagaoka},  where a similar approach 
to \cite{VerduHan_general} is made for the general 
(classical-)quantum channels.  As an application 
the capacity formula for quantum stationary memoryless 
channels \cite{Holevo, Schumacher-Westmoreland}
is provided with a new simple proof by linking it to 
the quantum Stein's lemma via our Theorem~\ref{Thm:D=infD=supD}. 

Finally, we mention some remarkable progresses in related subjects 
reported after submitting the accepted version of the present paper.
The quantum  Chernoff bound for 
symmetric Bayesian discrimination of two i.i.d.\ 
states has been established by \cite{chernoff_lower} and \cite{chernoff_upper}. 
Based on an inequality shown in  \cite{chernoff_upper}, 
it has been proved by \cite{Hayashi_hoeffding} that 
$B_e (r\,|\,\rhovec\,\|\,\sigmavec )$ for the quantum i.i.d.\ 
case satisfies (cf.\ equation~(\ref{eq:hoeffding}))
\[
B_e (r\,|\,\rhovec\,\|\,\sigmavec )
\geq \max_{-1\leq\theta <0}\frac{(1+\theta)r +\psi(\theta)}{\theta} ,
\]
where $\psi(\theta)\defeq \log\Tr\rho^{1+\theta}\sigma^{-\theta}$. 
This is the tightest lower bound on $B_e (r\,|\,\rhovec\,\|\,\sigmavec )$ 
of those obtained so far.  Moreover it seems natural to conjecture that 
the bound achieves the equality in general.

\begin{biography}{Hiroshi Nagaoka (M'88)} was born in Tokyo in 1955.  He received 
the B. Eng. and M. Eng. degrees in 1980 and 1982, respectively, 
from the University of Tokyo and the Dr. Eng. degree from 
Osaka University in 1987.  

He worked at the Tokyo Engineering University from 
1986 to 1989, at Hokkaido University from 1989 to 1993, 
and is currently an associate professor at the University of 
Electro-Communications.  His research interests include 
information geometry, quantum information theory and 
quantum statistical inference.  

\end{biography}

\begin{biography}{Masahito Hayashi} was born in Japan in 1971.
He received the B. S. degree from Faculty of Sciences in Kyoto 
University, Japan, in 1994
and the M. S. and Ph. D. degrees in Mathematics from 
Kyoto University, Japan, in 1996 and 1999, respectively.

He worked in Kyoto University as a Research Fellow of the Japan Society of the 
Promotion of Science (JSPS) from 1998 to 2000,
and worked in the 
Laboratory for Mathematical Neuroscience, 
Brain Science Institute, RIKEN from 2000 to 2003.
In 2003, he joined 
Quantum Computation and Information Project, 
Japan Science and Technology Agency (JST)
as the Research Head.
He also works in 
Superrobust Computation Project
Information Science and Technology Strategic Core (21st Century COE by MEXT)
Graduate School of Information Science and Technology
The University of Tokyo
as Adjunct Associate Professor from 2004.
He is an Editorial Board of International Journal of Quantum Information.
His research interests include quantum information theory and
quantum statistical inference.
\end{biography}


\begin{thebibliography}{99}
\bibitem{New}
M. Hayashi,
``Second order asymptotics in fixed-length source coding and intrinsic 
randomness,''
LANL E-print cs.IT/0503089.
\end{thebibliography}
\end{document}